\documentclass[12pt]{article}
\usepackage{amsmath}
\usepackage{natbib}
\usepackage{algorithm}
\usepackage{algpseudocode}
\usepackage{lscape}
\usepackage{setspace}
\usepackage{authblk}
\usepackage{graphics,graphicx}
\usepackage{appendix}
\usepackage{wrapfig}
\usepackage{caption}
\usepackage{subcaption}
\usepackage{subfig}
\usepackage{url}
\usepackage{amsmath,amssymb,amsthm}
\usepackage{bm}
\usepackage{graphicx}
\usepackage{booktabs}
\usepackage{comment}
\usepackage{enumitem}
\usepackage[nomarkers,nolists]{endfloat}
\setlist{nolistsep}
\newtheorem{theorem}{Theorem}

\newtheorem{corollary}{Corollary}

\newtheorem{proposition}{Proposition}

\newcommand{\betavec}{\boldsymbol{\beta}}

\newcommand{\onevec}{\boldsymbol{1}}

\newcommand{\yvec}{\boldsymbol{y}}

\newcommand{\tr}{{\mathop{\mathrm{tr}}}}

\DeclareMathOperator*{\argmin}{arg\,min}
\DeclareUnicodeCharacter{0308}{HERE!HERE!}

\newcommand{\blind}{1}

\begin{document}

\def\spacingset#1{\renewcommand{\baselinestretch}%
{#1}\small\normalsize} \spacingset{1}


\if1\blind
{
  \title{\bf Large Row-Constrained Supersaturated Designs for High-throughput Screening}

    \author[1]{Byran J. Smucker\thanks{Corresponding author: Byran Smucker, with current affiliation Department of Public Health Sciences, Henry Ford Health, 1 Ford Place, Suite 3C, Detroit, MI 48202 (bsmucke1@hfhs.org)}
}
    \author[1]{Stephen E. Wright}
    \author[1]{Isaac Williams}
    \author[2]{Richard C. Page}
    \author[3]{Andor J. Kiss}
    \author[2]{Surendra Bikram Silwal}
    \author[4]{Maria Weese}
    \author[5]{David J. Edwards}
    
    \affil[1]{Department of Statistics, Miami University, Oxford, OH, 45056}
    \affil[2]{Department of Chemistry and Biochemistry, Miami University, Oxford, OH, 45056}
    \affil[3]{Center for Bioinformatics and Functional Genomics, Miami University, Oxford, OH, 45056}
    \affil[4]{Department of Information Systems \& Analytics, Miami University, Oxford, OH, 45056}
    \affil[5]{Department of Statistical Sciences and Operations Research, Virginia Commonwealth University, Richmond, VA, 23284}

  \maketitle
} \fi

\if0\blind
{
  \bigskip
  \bigskip
  \bigskip
  \begin{center}
    {\LARGE\bf Large Row-Constrained Supersaturated Designs for High-throughput Screening}
\end{center}
  \medskip
} \fi

\bigskip
\begin{abstract}
High-throughput screening, in which large numbers of compounds are traditionally studied one-at-a-time in multiwell plates against specific targets, is widely used across many areas of the biological sciences including  drug discovery.  
To improve the efficiency of these screens, we propose a new class of supersaturated designs that guide the construction of pools of compounds in each well. Because the size of the pools are typically limited by the particular application, the new designs accommodate this constraint and are part of a larger procedure that we call Constrained Row Screening or CRowS. We develop an efficient computational procedure to construct the CRowS designs, provide some initial lower bounds on the average squared off-diagonal values of their main-effects information matrix, and study the impact of the constraint on design quality. We also show via simulation that CRowS is statistically superior to the traditional one-compound-one-well approach as well as an existing pooling method, and demonstrate the use of the new methodology on a Verona Integron-encoded Metallo-$\beta$-lactamase 2 assay.
\end{abstract}

\noindent%
{\it Keywords:} experimental design; biological screening; the Lasso
\vfill

\newpage
\spacingset{1.9} 

\section{Introduction}\label{sec:introduction}

High-throughput screening (HTS) of chemical or biological compounds is used extensively across a range of critical scientific enterprises. This type of experiment is common in basic functional genomics studies and synthetic biology applications \citep{zeng2020high, sarnaik2020high}, but arguably its highest impact application is drug discovery \citep{blay2020high, soheilmoghaddam2021high,aldewachi2021high}. The literature suggests that nearly one-third of clinical drug candidates emerge from high-throughput screening campaigns \citep{brown2018recent}, and these screens were a centerpiece in the National Cancer Institute's Cancer Moonshot program \citep{wilson2020creating}. For example, HTS is used in the search for inhibitors of enzymes that cause antibiotic resistance. The prevalence of such enzymes is of great concern because they exist in bacteria that cause nosocomial infections, including in secondary bacterial infections contributing to an estimated 50\% of COVID-19 deaths \citep{Mojica2022-xg, Chedid2021-eo}. As liquid handling, miniaturization, and other screening technologies have improved, the number of potentially inhibitive compounds to explore has increased as well \citep{elkin2015just,volochnyuk2019evolution}, which motivates the need to improve the experimental and statistical aspects of these screens. 

Classically, a single compound is applied to a single well in a 96-, 384-, or 1536-well plate \citep{Malo2006, kainkaryam2009pooling}, an assay is performed with relevant normalization and bias-correction \citep{altman2005replication,birmingham2009statistical,malo2010experimental,dragiev2011systematic, caraus2015detecting}, and hit compounds are identified with a z-score. As an alternative to one-compound-one-well (OCOW), researchers have suggested creating pools of compounds in each well which can improve HTS because compounds are observed in multiple wells rather than just one. Various versions of pooling have been used successfully \citep[e.g.,][]{devlin1996high, wilson1996deconvolution,  motlekar2008evaluation, kainkaryam2009pooling, elkin2015just, silva2016output}, and the most aggressive version, called non-adaptive pooling \citep{thierry2006new,kainkaryam2009pooling}, is said to accurately screen large numbers of compounds with relatively few pools. However, these methods are largely non-statistical in both their design and analysis. 

In this work, we develop a novel class of row-constrained supersaturated designs that provides the pooling instructions to run a $k$-compound screen in $n$ wells, while accounting for a screener-specified constraint on the maximum number of compounds per well. The constraint may be due, for instance, to the size of the well, a required minimum compound concentration, or the maximum ionic strength of the solution beyond which the compounds may precipitate. We call these limitations \emph{row constraints}, denoted by $c$, because they limit the number of $+1$'s in each design row. Designs are constructed using the unconditional $E(s^2)$ criterion from the supersaturated design literature, and we analyze them using the Lasso. These designs are part of a larger procedure, which we call Constrained Row Screening, or CRowS, that includes both the design and its analysis via the Lasso.

We demonstrate that altogether CRowS is more flexible, cost-effective, and statistically efficient than traditional methods. We present a computationally efficient heuristic to construct the designs of any relevant $(n,k,c)$ combination, study the designs both theoretically and empirically as a function of $c$, provide evidence that our methodology improves upon both the OCOW approach as well as the pooling method of \citet{kainkaryam2008poolhits}, and show in a set of real experiments that it can be used to identify an inhibitor of a particular antibiotic-resistant enzyme. Thus, our contributions are a statistically principled and improved solution to an important biological problem and innovation in the construction of the relevant designs.  (We note that after our work was completed we became aware of a recent paper by \citet{ji2023target} that constructs pools by optimizing a ``sensing matrix"  which also uses a version of the Lasso to identify hits. They don't explicitly constrain the size of the pools, nor do they engage with the statistical literature or study their method theoretically or via simulation. Still, it is encouraging that the biomedical community is considering statistical approaches to solve high-throughput screening problems.)

Our work emerges from research on supersaturated design (SSD), where $k$ factors are investigated using $n$ runs (with $n\leq k$). SSDs, which were first described over sixty years ago \citep{satterthwaite1959random,booth1962some}, have seen extensive development over the last thirty years \citep[e.g.,][]{lin1993new,wu1993construction,nguyen1996algorithmic,cheng1997s, georgiou2014supersaturated,jones2014optimal}. In our setting, an SSD is a two-level, $n \times k$ design $X$, composed of $k$ compounds (factors; absence with level $-1$; presence with level $+1$), and $n$ wells (runs) with $k \geq n$ (along with a constraint on the design rows, described below and in Section \ref{sec:r-cSSD}). The SSD provides a set of instructions regarding which compounds are included in which well, so one may improve the design by constructing it using a statistically relevant criterion. SSDs are classically constructed using the $E(s^2)$ criterion \citep{booth1962some, nguyen1996algorithmic} to create balanced columns, as close to mutually orthogonal as possible. The row constraint of Section \ref{sec:r-cSSD} typically necessitates a design with more $-1$'s than $+1$'s, so we use the unconditional $E(s^2)$ criterion \citep{marley2010comparison, jones2014optimal,weese2015searching} which has the same orthogonality goal without the requirement of column balance. Mathematically, let $L=[\mathbf{1},X]$, and $S=L^{'}L$ with elements $s_{ij}$ and 
$O_{S}=\sum_{i<j}s_{ij}^{2}$. Then the unconditional $E(s^2)$ criterion is

\vspace{-40pt}
\begin{align}
    UE(s^2) = \frac{O_{S}}{k(k+1)}, \label{eq:uessq}
\end{align}
\vspace{-40pt}

\noindent the average of the squared off-diagonal elements of $S$. In many cases $UE(s^2)$-optimal designs can be constructed directly \citep{jones2014optimal}, but for our row-constrained versions, we construct the designs algorithmically; see Section \ref{sec:alg}. After the design is constructed and executed, it is analyzed assuming the main-effects model 

\vspace{-50pt}
\begin{align}
    \mathbf{y} = \beta_{0}\mathbf{1} + X\boldsymbol\beta + \boldsymbol\epsilon, \label{eq:model}
\end{align}
\vspace{-50pt}

\noindent where 
$\boldsymbol\beta=(\beta_1,\dots,\beta_k)^T$ and $\boldsymbol\epsilon \sim N(\mathbf{0},\sigma^2 I)$. Because $k\geq n$, ordinary least squares cannot be used to estimate factor effects and 
regularization methods are preferred for analysis. The Gauss-Dantzig selector \citep{candes_tao2007} is effective \citep{phoa2009analysis, marley2010comparison, weese2015searching}, while the closely related Lasso \citep{tibshirani1996regression, meinshausen2007discussion, lounici2008sup, james2009dasso, asif2010lasso} achieves similar performance \citep[see][]{draguljic2014screening}.  \citet{stallrich2023optimal} has also recently developed an optimal screening design framework based on the Lasso. Thus, to analyze our proposed designs we use the Lasso, which selects estimates according to 

\vspace{-40pt}
\begin{align}
(\hat{\beta}_0,\hat{\betavec})=\argmin_{\beta_0,\betavec} \frac{1}{2n}||\yvec-\beta_0\onevec-X\betavec||_2^2 + \lambda ||\betavec||_1, \label{eq:lasso}   
\end{align} 
\vspace{-40pt}

\noindent where $||\betavec||_\ell=(\sum_{j=0}^{k} |\beta_j|^\ell)^{1/\ell}$ and $\lambda >0$.

The remainder of the paper is structured as follows. In Section \ref{sec:r-cSSD}, we describe CRowS designs in more detail, along with the algorithm we use to construct them. Section \ref{sec:investigation} provides initial lower bounds on the $UE(s^2)$ values for these CRowS designs, as well as an empirical study that sheds light on their behavior as a function of the severity of the row constraint. In Section \ref{sec:sim} we use a set of simulations to show the effectiveness of our proposed procedure compared to alternative methods. In Section \ref{sec:real} we provide more detail on the metallo-$\beta$-lactamase system, and describe the results of several screens as a real-world proof-of-concept of CRowS. We provide a discussion and conclusion in Section \ref{sec:discussion}.

\section{Row-Constrained Supersaturated Designs} \label{sec:r-cSSD}

In this section we describe the new class of supersaturated designs that we propose, along with the heuristic optimization procedure developed to construct them. As explained in Section \ref{sec:introduction}, $UE(s^2)$-optimal designs minimize the average of the squared off-diagonal elements of information matrix $S$. In the present setting, we propose a generalization of $UE(s^2)$-optimal designs that impose a constraint on the number of $+1$'s in each row. Thus, using notation from Section \ref{sec:introduction}, we say that an $n \times k$ design $X$ is a CRowS design if $n \leq k$, $[X]_{ij}=x_{ij}$, and the design solves the following optimization problem:

\vspace{-40pt}
\begin{align}
    \text{min} \;\; &UE(s^2) \\
    \text{s.t.} \;\; &x_{ij} \in \{-1,1\} \; \forall i,j \\
    & \sum_{j=1}^{k} x_{ij} \leq 2c-k \; \forall i \label{eq:c-constr}
\end{align}
\vspace{-40pt}

\noindent where $UE(s^2)$ is defined in \eqref{eq:uessq} and $c$ is the maximum number of compounds/well.

\subsection{Construction Algorithm} \label{sec:alg}

In order to construct approximately optimal row-constrained $UE(s^2)$ designs, we use the coordinate exchange (CEx) heuristic \citep{meyer_nachtsheim1995}, along with some computational devices to make the procedure faster. Let $n_{1i}$ be the number of $+1$'s in design row $x_{i}$, so that $n_{1i} \leq c \; \forall i$. The algorithm construction is described in Algorithm \ref{alg:overall}. For the designs in this paper, we ran Algorithm \ref{alg:overall} with 100 random starts and chose the design with the best objective function value.

\begin{algorithm} 
    \caption{Row-Constrained Design Construction}
    \label{alg:overall}
    \begin{algorithmic}[1]
        \State Randomly generate an initial design with $n_{1i} \leq c$ for all rows $i$.
        \State For each row $i=1,\ldots,n$ in order: \label{alg-code:outerloop}
  \begin{enumerate}
   \item[a.] [1-CEx] For each column $j=1,\ldots,k$ in order: if
       $x_{ij}=+1$ or $n_{1,i}<c$, then change the sign of $x_{ij}$ if
       doing so improves the criterion.
       \label{alg-code:1coord-flipeither}
   \item[b.] [2-CEx] Let $\mathcal H$ denote $\{j:x_{ij}=+1\}$ and let
       $\mathcal L$ denote $\{j:x_{ij}=-1\}$ using the current values in
       $X$. For each element $j\in\mathcal H$ in order, do the following:
       \label{alg-code:2-coord}
    \begin{enumerate}
     \item[] Find the column index $l\in\mathcal L$ for which swapping
         the signs of $x_{ij}$ and $x_{il}$ most improves the
         criterion. If such an improvement is possible, then swap the
         signs of $x_{ij}$ and $x_{il}$ and also replace $\mathcal L$
         with $(\mathcal L\cup\{j\})\setminus\{l\}$.
    \end{enumerate}
  \end{enumerate}    
  \State Repeat step \ref{alg-code:outerloop} until no further exchanges can be made.
  \end{algorithmic}
    
\end{algorithm}

\subsection{Computational Improvements} \label{sec:comp_improvements}

\citet{weese2017criterion} discuss how to quickly update the $E(s^{2})$ criterion when changing a coordinate in the design. Here we use a similar strategy while assuming $n_{1i}\leq c$. In steps 2a and 2b of the algorithm, we test if an exchange under consideration improves the criterion. If it does, then we update the design matrix, information matrix, and design criterion value accordingly. These updates are highly amenable to an implementation using vectorized operations, and our implementation also attempts to minimize the number of operations needed for
each such test and update. 

As before, we let $L=[\mathbf1,X]$ and $S=L^{'}L$. Defining $Q(X) := \tr(L^{'}LL^{'}L) = \tr(S^2)$, we may write $UE(s^2) = [Q(X)-n^{2}(k+1)]/(k(k+1))$. (Throughout this section and the next, we use the notation $A_{:,j}$ to represent the $j$\textsuperscript{th} column of a matrix $A$.) We need to determine how much $Q(X)$ changes as the result of either a single-coordinate
or two-coordinate exchange in $X$. For notational convenience in indexing, we
use $x_{i0}=1$ to denote entries in the first column of $L$. This allows us
to write
\[ Q(X) = \sum_{j=0}^k\sum_{l=0}^k s_{jl}^2
   =  \sum_{j=0}^k\sum_{l=0}^k \bigg(\sum_{h=1}^nx_{hj}x_{hl}\bigg)^2
\]
Our test-and-update procedure is based on the following observation.

\begin{proposition}
Consider two design matrices $X,\tilde X\in\{\pm1\}^{n\times k}$ with
$L=[\mathbf1,X]$ and $S=L'L$. If $X$ and $\tilde X$ differ only in row $i$,
then for $J:=\{j=0,\ldots,k \mid \tilde x_{ij}=-x_{ij} \}$ we have
\[ Q(\tilde X)-Q(X)
   =8|J|(k+1-|J|)
    -8\sum_{j\in J}\bigg[x_{ij}\sum_{l\not\in J}s_{jl}x_{il}\bigg].
\]
\end{proposition}

\noindent The proof of this result is given in Section 1 of Supplementary Material A. The proposition provides a simple test to determine if switching the signs of all
$x_{ij}$ for fixed $i$ and $j\in J$ improves the criterion. The following
result specializes it to the 1-coordinate and 2-coordinate exchanges in the
proposed algorithm.

\begin{corollary} \label{cor:coex}
Consider a design matrix $X\in\{\pm1\}^{n\times k}$ with $L=[\mathbf1,X]$ and
$S=L'L$ and let $L_{i[j]}$ denote the row vector given by replacing element
$j$ in the row vector $L_{i:}$ with a zero. (Recall that the columns of $L$
are indexed from 0.)
\begin{enumerate}
\item[a.] Swapping signs on $x_{ij}$ decreases $Q(X)$ if and only if
    $k<x_{ij}L_{i[j]}S_{:,j}$. In this case, the change in $Q(X)$ will be
    $8(k-x_{ij}L_{i[j]}S_{:,j})$ and the corresponding rank-2 update of $S$
    is given by subtracting $2x_{ij}L_{i[j]}$ from row $j$ of $S$ and
    subtracting its transpose from column $j$.
\item[b.] Suppose $x_{ij}=+1$ and $x_{il}=-1$. Swapping signs on both
    $x_{ij}$ and $x_{il}$ decreases $Q(X)$ if and only if $2(k-1)+n <
    L_{i[j]}S_{:,j}-L_{i[j]}S_{:,l}+s_{jl}$. Among all such indices $l$, the
    improvement in the criterion will be greatest for those minimizing
    $L_{i[j]}S_{:,l}-s_{jl}$. In this case, the change in $Q(X)$ will be
    $8[2(k-1)+n-L_{i[j]}S_{:,j}+L_{i[j]}S_{:,l}-s_{jl}]$. We update $S$ by
    subtracting $2x_{ij}L_{i[j]}$ from row $j$ of $S$, adding
    $2x_{il}L_{i[l]}$ to row $l$, and also applying the corresponding
    column updates.
\end{enumerate}
\end{corollary}

Notice that Corollary \ref{cor:coex} allows us to test for decrease without updating $S$.
It also shows that we can vectorize the $l$-index selection for 2-coordinate exchange by calculating the row vector $L_{i[j]}S_{:,\mathcal
L}+S_{j,\mathcal L}$ and finding its minimum entry.

\section{Properties of CRowS Designs} \label{sec:investigation}

In this section, we study the designs described in Section \ref{sec:r-cSSD}. In particular, we provide some initial bounds on the $UE(s^2)$ criterion for the row-constrained designs, under the assumption that the row constraints are tight, which provides a design optimality check. Then, we empirically investigate several relevant design settings to better understand how the designs change as a function of the row constraint, providing insights into the practical effect of the constraints on the effectiveness of the row-constrained designs.

\subsection{Some Bounds for CRowS Designs} \label{sec:bounds}

The result in this section uses ideas adapted from \citet{liu2002s}; it is proven in Section 2 of Supplementary Material A. Recall the notation $Q(X) := \tr(L^{'}LL^{'}L)$ and $L=[\mathbf1,X]$ from 
\S\ref{sec:comp_improvements}. It is easily verified that $Q(X) = n^2 + 2\|X'\mathbf1_n\|_2^2 + \tr(X'XX'X)$. 

\begin{theorem}
We have
\begin{equation}
 Q(X)
 = n^2(1-k^2) + 2\|X'\mathbf{1}_n\|_2^2 + 2n\|X\mathbf1_k\|_2^2
   + \sum_{l=1}^k\sum_{j=1}^k\|X_{:,l}-X_{:,j}\|_1^2.
 \label{eq:Qpwquadratic}
\end{equation}
If $X\mathbf{1}_k=(2c-k)\mathbf{1}_n$ then the terms in
(\ref{eq:Qpwquadratic}) satisfy
\begin{gather}
 \|X\mathbf1_k\|_2^2=n(2c-k)^2, \label{eq:rowsumnrmsq} \\
 \|X'\mathbf1_n\|^2_2 \geq
 (k-\delta)(n-2\gamma)^2 + \delta(n-2\gamma-2)^2,
  \label{bnd:colsumnrmsq} \\
 \sum_{l=1}^k\sum_{j=1}^k\|X_{:,l}-X_{:,j}\|_1^2\geq
   4[(k^2-k)\phi^2 + \psi(2\phi+1)]
   \label{bnd:rowsumcoldiff}
\end{gather}
for integers $\gamma := \lfloor nc/k \rfloor$, $\delta := nc-k\gamma$, $\phi
:= \lfloor 2nc(k-c)/(k^2-k) \rfloor$, and $\psi := 2nc(k-c) -(k^2-k)\phi$.
\end{theorem}

If the bound (\ref{bnd:colsumnrmsq}) is attained, then all the column sums for $X$ take values from $\{\gamma,\gamma+1\}$ in specific proportions. Likewise, if the bound (\ref{bnd:rowsumcoldiff}) is attained, then all $l^1$-differences of columns in $X$ take values from $\{2\phi,2\phi+2\}$. That raises several questions. When must there exist feasible $X$ with all column sums in $\{\gamma,\gamma+1\}$? Must there exist feasible $X$ with all $\|X_{:,l}-X_{:,j}\|_1$ in $\{2\phi,2\phi+2\}$? Can both be satisfied simultaneously? We leave the study of these questions to future work.

\subsection{Empirical Investigation of Row Constraint} \label{sec:c_study}

An additional unanswered question is whether there is a threshold for $c$ below which we are guaranteed that a row-constrained $UE(s^2)$-optimal design will have tight row constraints. In this section we investigate this question empirically. Reasoning intuitively, unconstrained $UE(s^2)$-optimal designs try to make $s_{jl}$'s small, which means that the columns are pushed toward balance in order that the $s_{0j}$'s are small. In an unconstrained design, we therefore expect $+1$'s and $-1$'s should each occur approximately $nk/2$ times. Thus, for $c\geq k/2$, it is possible for the numbers of $+1$'s and $-1$'s to occur the same number of times, and we may observe slack in constraint \eqref{eq:c-constr} with the optimal $UE(s^2)$ value similar to the unconstrained case. For $c<k/2$, however, we may see tight row constraints and, as $c$ gets smaller, an increasingly large difference between the row-constrained optimal $UE(s^2)$ value and that of the unconstrained optimal design. That is, if $U^{*}$ is the $UE(s^2)$ value for the unconstrained $UE(s^2)$-optimal design and $U_{c}^{*}$ is the $UE(s^2)$ value of the row-constrained $UE(s^2)$-optimal design, then $U^{*}\leq U_{c}^{*}$ if $c \leq k/2$, but as $c \rightarrow k$, $U_{c}^{*} \rightarrow U^{*}$. 

To explore the relationship between $c$ and $UE(s^2)$, along with the effect of $c$ on the tightness of the constraints in \eqref{eq:c-constr}, we first examine a set of approximately $UE(s^2)$-optimal designs with $n=96$, $k=144$ and $c\in \{2,3,\ldots, 144\}$, each constructed using $100$ algorithm tries of the algorithm described in Section \ref{sec:alg}. We consider several properties of the designs, each as a function of $c$: the $\text{UE}(s^2)$ criterion value, along with 
the minimum, mean, and maximum slack in the row constraint, where slack is measured for each row $i$ as $(2c-k)-\sum_{j=1}^{k}x_{ij}$. We also plot the bounds from Section \ref{sec:bounds} for those designs without row constraint slack, and provide some simulation results. For the simulations, we follow the approach we describe in Section \ref{sec:sim}, assuming $a=1$ nonzero element of $\boldsymbol\beta$ with value $\beta$, in line with application-specific sparsity levels of 1\% or less. Since we know the desired effect directions, we consider $\beta \in \{0.375,0.5,0.75,1\}$, which translate to differences in mean response (pools with vs. pool without active compound) of $0.75$, $1$, $1.5$, and $2$, respectively. The datasets are analyzed using the Lasso as in Section \ref{sec:sim-CRowS}. 

\begin{figure}[h]
\centering
\includegraphics[width=160mm]{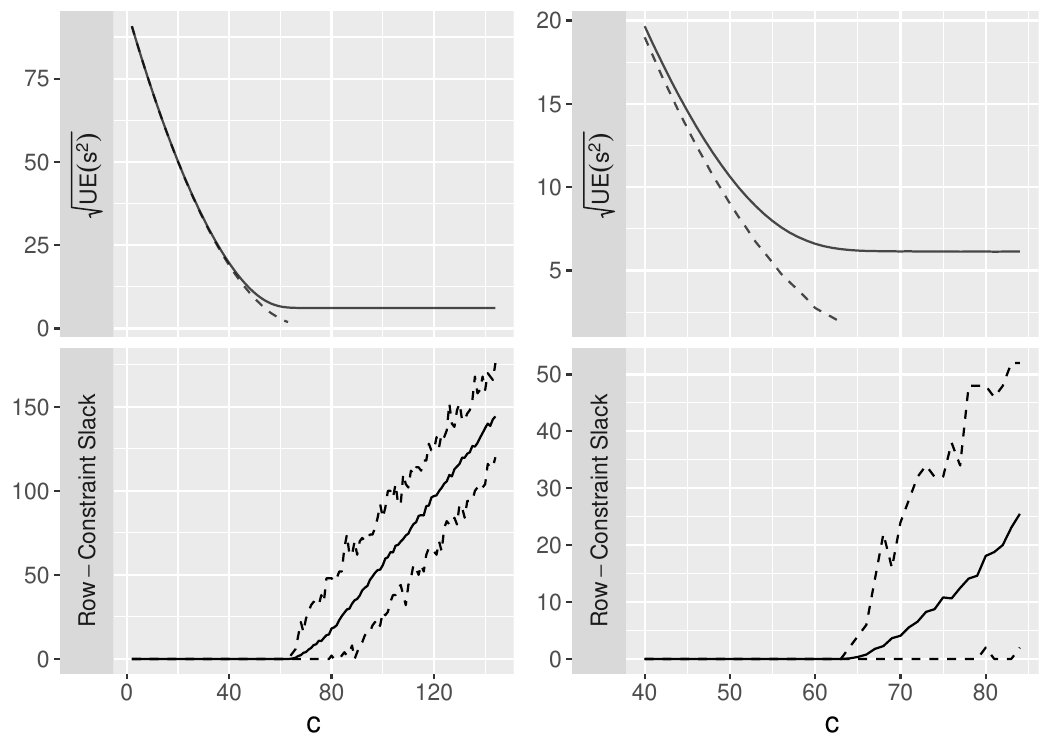}
\caption{Characteristics of CRowS designs with $n=96$, $k=144$ and $c \in \{2,3, \ldots, 144\}$. Plots are a function of $c$, the number of $+1$'s allowed in each row of design. Top left: $UE(s^2)$ value for each design (solid line) along with the bounds from Section \ref{sec:bounds} (for all designs with no row-constraint slack) (dashed line). Bottom left: $(2c-k)-\sum_{j=1}^{k}x_{ij}$; the solid line is the average slack, while the lower and upper dashed lines are the minimum and maximum slack. Top right: The same as the top left plot, for $40 \leq c \leq 84$. Bottom right: The same as the bottom left plot, for $40 \leq c \leq 84$.}
\label{fig:n96_k144_cs}
\end{figure}

\begin{figure}[h]
\centering
\includegraphics[width=130mm]{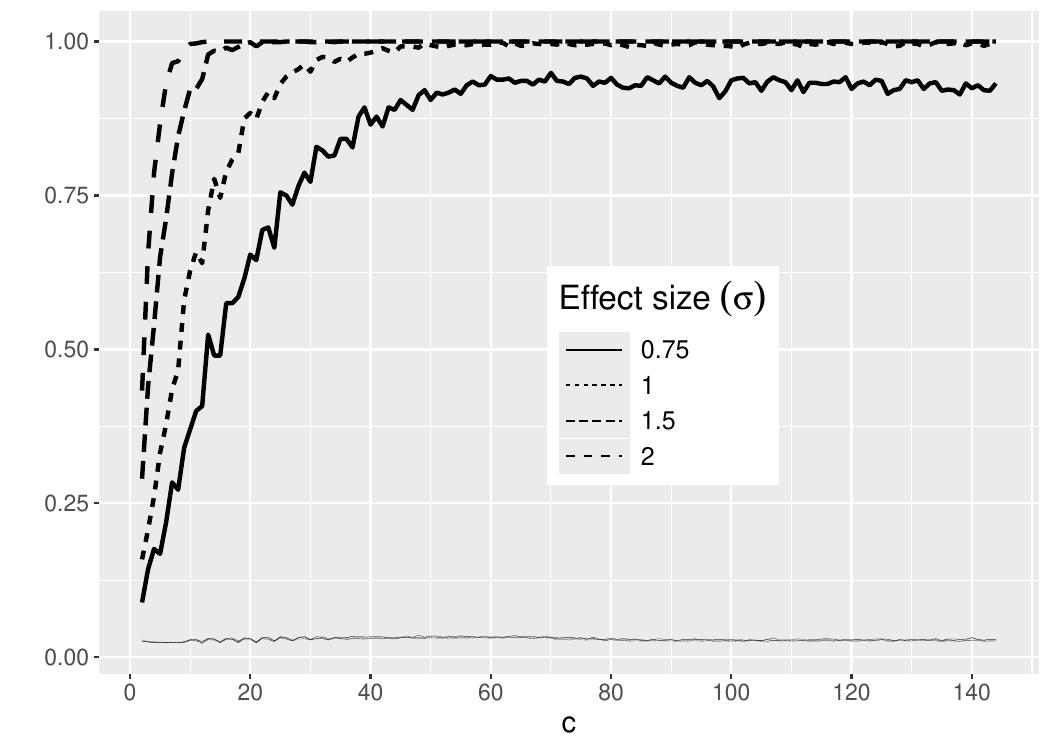}
\caption{Simulation results for CRowS designs with $n=96$, $k=144$ and $c \in \{2,3, \ldots, 144\}$, for four effect sizes, assuming one active factor; the heavier lines above are true positive rates while the lighter lines below are false positive rates.}
\label{fig:sim_plot_cs}
\end{figure}

Figure \ref{fig:n96_k144_cs} provides a visualization of $\sqrt{UE(s^2)}$ and the measures of row-constraint slack for all $c$ (left) and for a smaller set of $c$ (right). In Figure \ref{fig:sim_plot_cs}, we provide simulation results as a function of $c$. We observe that between $c=60$ and $c=70$ the constraint ceases to measurably impact the criterion value, the amount of row-constraint slack, and the TPR. We also see that while the bounds appear to be fairly tight (Figure \ref{fig:n96_k144_cs}, top left), closer inspection (Figure \ref{fig:n96_k144_cs}, top right) shows that there is a substantial gap as $c$ approaches the value beyond which there will be row-constraint slack. Thus, we conclude that for these designs any row constraint larger than $k/2=72$ will not have an appreciable effect on the results of an experiment using them, though the constraint includes slack beginning at some value $c$ slightly smaller than $k/2$. We have seen the same pattern for a number of other design sets as well (see Section 3 of Supplementary Material A). The simulation (Figure \ref{fig:sim_plot_cs}) also shows that even for $c$ as small as 5 or 10, the designs reliably categorize active and inactive factors for sufficiently large effect sizes. 

\section{Comparison of Biological Screening Methods} \label{sec:sim}

Most commonly, biological screening is conducted using a simplistic one-compound-one-well (OCOW) approach \citep{Malo2006, kainkaryam2009pooling, malo2010experimental}. Alternatively, a non-statistical pooling strategy has been proposed \citep{kainkaryam2008poolhits} (poolHiTS). In this section, we compare via simulation these existing methods with CRowS, our proposed procedure, which uses the designs of Section \ref{sec:r-cSSD}, along with the Lasso, to estimate the set of hit compounds. We have compared the approaches with regard to their statistical performance, measuring both the true positive rate (TPR, proportion of truly active factors identified as active) and false positive rate (FPR, proportion of truly inactive factors declared active).

\subsection{Methods to Compare} \label{sec:methods}

For the sake of the comparative results in this section, we make several simplifying assumptions. First, we assume that we know the effect directions of interest. For instance, in our metallo-$\beta$-lactamase application, the goal is the inhibition of enzyme activity and so we look for a negative effect direction. Without loss of generality, we assume a positive effect direction in our simulations. Secondly, we assume a known ambient level of variability, $\sigma^{2}$, from model \eqref{eq:model}. This is a strong assumption but simplifies the analyses, particularly for OCOW and poolHiTS. Third, we assume a known background mean response, $\mu$, when there are no hits in a particular well. (We provide an alternative set of simulations in Section 5 of Supplementary Material A that relaxes the assumptions about $\mu$ and $\sigma^{2}$.) Finally, we assume that responses, or some standard transformation of them, can be reasonably approximated by a normal distribution.

\subsubsection{One-Compound-One-Well (OCOW)}

The first approach is the most straightforward: apply a single compound to a single well and evaluate whether the result is unexpected in the assumed effect direction. We are assuming normal responses, a known positive effect direction, and known values of $\sigma^{2}$ and $\mu$. So the analysis is quite simple: declare as active any compound whose response exceeds $\mu + F^{-1}(0.95)\sigma$, where $F^{-1}$ is the inverse normal distribution function.

\subsubsection{poolHiTS} \label{sec:poolHiTS}

The poolHiTS procedure, using Shifted Transversal Designs \citep[STD,][]{kainkaryam2008poolhits}, is a pooling method which also provides for row constraints. Its designs are intuitive and easy to construct, but not based upon a statistically motivated criterion. Similarly, its analysis procedure is not based upon the estimation of a statistical model. These designs also are unavailable for many $(n,k,c)$ combinations. In our comparisons in Section \ref{sec:sim}, we had to choose $(n,k,c)$ values that STD allowed, rather than the most natural values. 
This methodology also assumes a binary response vector (that is, that each well can be categorized as a ``hit'' if it contains an active compound, or a ``miss'' if it doesn't), so in the standard numeric-response setting it is required to determine a threshold beyond which a well's response is categorized as a ``hit''. We conducted some tests to establish an appropriate threshold which would best balance TPR and FPR. In the end, assuming a positive effect direction, we chose $\mu + F^{-1}(0.96)\sigma$, where $F^{-1}$ is again the inverse standard normal distribution function. A description of this method is provided in Section 4 of Supplementary Material A.

\subsubsection{Constrained Row Screening} \label{sec:sim-CRowS}

Our proposed approach uses a CRowS design from Section \ref{sec:r-cSSD}, constructed using 100 algorithm randomized starts of the procedure described in Section \ref{sec:alg}, and then analyzed using the Lasso. For the simulations, where using a profile plot to choose active effects is impractical, we use a thresholding procedure to reduce FPR, with the threshold chosen via informal testing. In particular, our analysis procedure is the following: 
\begin{enumerate}[label=(\alph*)]
    \item Center and scale the design $X$ so that each column has the same length, call it $X_{cs}$; and center response $\mathbf{y}$, call it $\mathbf{y}_{c}$.
    \item Obtain Lasso solutions for $\log(\lambda)$ ranging from $-8$ to $\log(\max(|X_{cs}^{T}\mathbf{y}_{c}|))$, the range chosen to reasonably ensure a full exploration of the Lasso estimation profile.
    \item Across all $\lambda$, set any parameter estimate to 0 if it is smaller than $\sigma/8$. Because we assume a known effect direction, we threshold all negative estimates.
    \item Refit each model, across $\lambda$'s, using least squares, and choose the model with the smallest BIC. The factors with nonzero estimates for this model are declared hits.
\end{enumerate}  

\subsection{Description of Simulation Protocol}

For the two pooling procedures (STD and SSD), data are simulated from the model in Eq. \eqref{eq:model}, assuming without loss of generality that $\sigma^{2}=1$. For the designs $X$ in the model, we used a set of nine different designs, with $(n,k,c)$ combinations as given in Table \ref{table:1}. To more closely mimic conditions in a lab setting, we preferred to use a different set of designs---all nine combinations of ($n=96, k \in \{96,150,192\}, c \in \{10,30,50\}$)---but since STDs of these preferred sizes cannot be constructed, we accommodated the poolHiTS procedure by using the designs in Table \ref{table:1}. For the effect sizes $\boldsymbol\beta$, we considered effect sizes, in units of $\sigma$, of $D \in \{0.75, 1, 1.5, 2, 2.25, 3, 4\}$; that is, $D$ gives the difference in mean response between a pool which includes a hit compound and one that does not. Since our designs are parameterized such that $-1$ and $1$ represent the two levels, the elements of $\boldsymbol\beta$ are either $0$ or $D/2$. Also the parameterization leads to $\beta_{0}=\mu+a\times \frac{D}{2}$, where $\mu$ is the background response mean and $a$ and the number of active factors. We fix $a=1$ for all simulations in this section, as this represents between $0.5\%$ and $1\%$ of total compounds studied.

\begin{table}[h!]
\begin{center}
\caption{Design sizes for simulations. Both STD and CRowS designs  were constructed for each of the sizes.}
\begin{tabular}{ |c c c|c c c|c c c| } 
 \hline
  n & k & c & n & k & c & n & k & c \\
 \hline
 88 & 96 & 10 & 85 & 150 & 10 & 92 & 192 & 10 \\ 
 88 & 96 & 30 & 91 & 150 & 30 & 99 & 192 & 30 \\ 
 88 & 96 & 50 & 91 & 150 & 50 & 99 & 192 & 50 \\
 \hline
\end{tabular}
\label{table:1}
\end{center}
\end{table}

\subsection{Comparison of CRowS with Alternatives}

Here we provide a comparison of the three methods of Section \ref{sec:methods}, labeled OCOW, poolHiTS, and CRowS, in Figure \ref{fig:comp}. Since we are assuming known $\sigma^{2}$, the OCOW results will be unchanged across the nine design scenarios we consider up to variation across simulations, so we have only displayed it for the most difficult setting, ($n=96,k=192,c=10$).

In Figure \ref{fig:comp_tpr}, we clearly see the advantage of CRowS in terms of its ability to identify active compounds, compared to OCOW and poolHiTS. We also notice that the most restrictive constraint setting ($c=10$) results in a clear reduction in TPR, compared to less restrictive $c$ values. Furthermore, Figure \ref{fig:comp_fpr} shows that poolHiTS fails to control FPR for most of the settings simulated. In highly sparse systems where thousands of compounds are studied, large FPRs could mean dozens or even hundreds of extra follow-up experiments on ultimately inert factors.

\begin{figure}[h]
    \centering
     \begin{subfigure}{0.68\textwidth}
         \includegraphics[width=\textwidth]{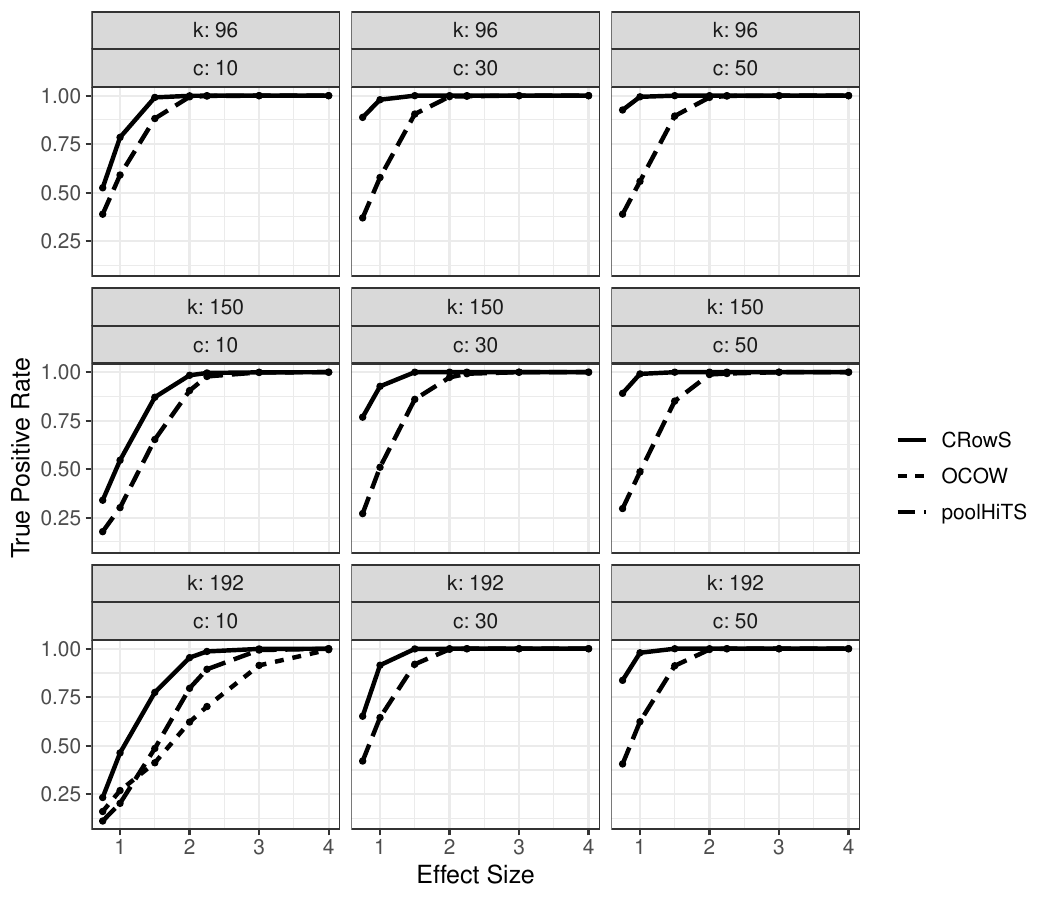}
         \caption{\small True Positive Rates for the three methods.}
         \label{fig:comp_tpr}
     \end{subfigure}
     \hfill
     \begin{subfigure}{0.68\textwidth}
         \includegraphics[width=\textwidth]{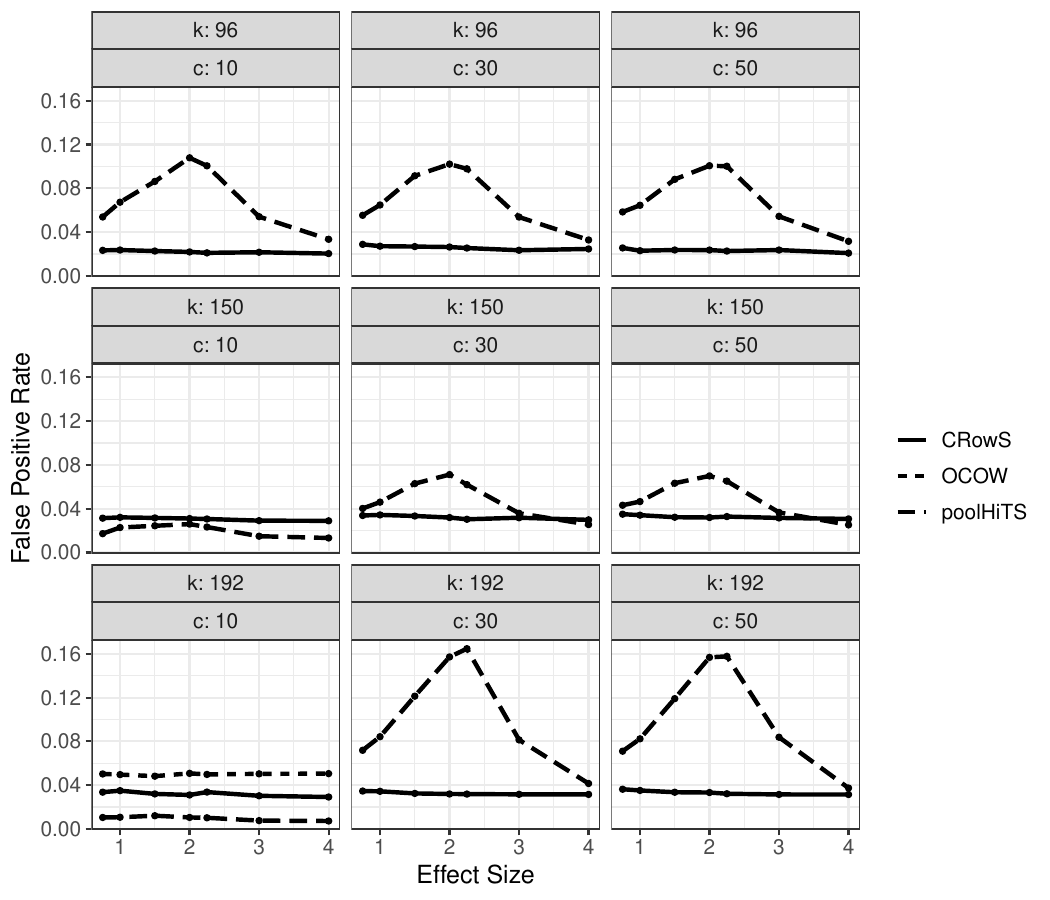}
         \caption{\small False Positive Rates for the three methods.}
         \label{fig:comp_fpr}
     \end{subfigure}
     \caption{\small Comparisons of three design/analysis approaches to biological screening: One-Compound-One-Well (OCOW), the proposed method (CRowS), and the poolHiTS procedure with shifted transversal designs (poolHiTS). For CRowS and poolHiTS, the number of wells are specified in Table \ref{table:1}; for OCOW, the number of wells is equal to $k$.}
     \label{fig:comp}
\end{figure}

\section{Real Experiments} \label{sec:real}

\begin{figure}
\centering
\includegraphics[width=162mm]{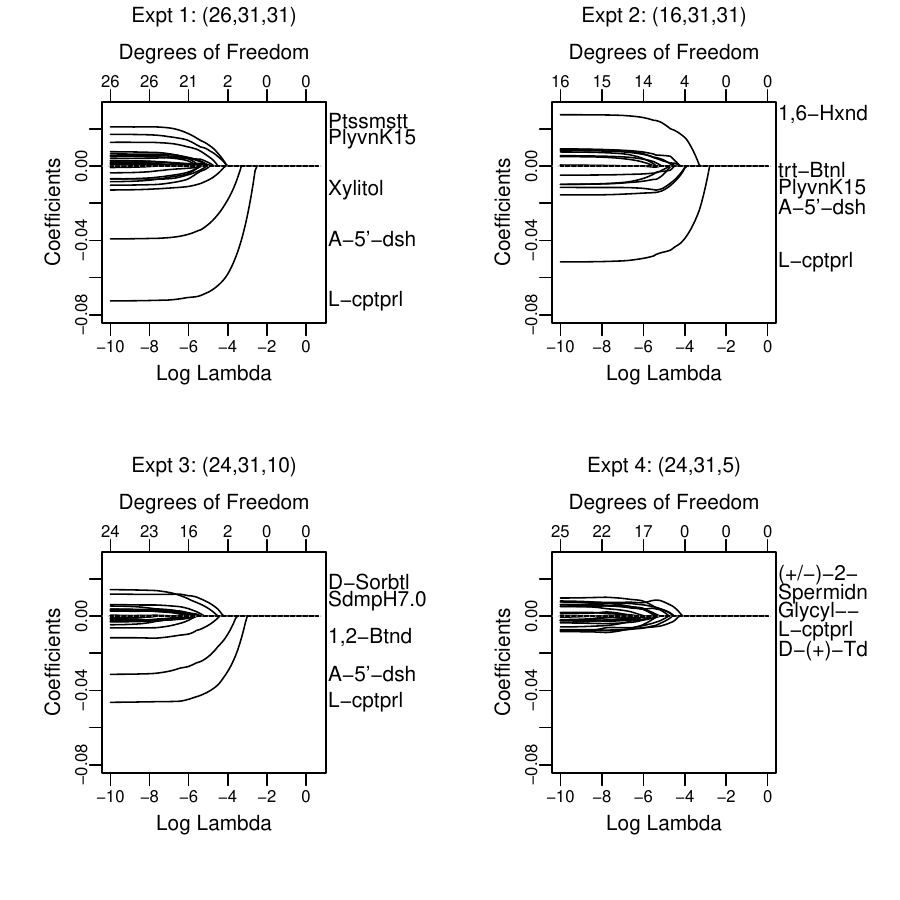}
\caption{Profile plots for four proof-of-concept experiments. For each plot, the largest five effects are annotated. The experiments are identified by $(n,k,c)$.}
\label{fig:expts}
\end{figure}

Metallo-$\beta$-lactamases exist in bacteria that cause nosocomial infections, including secondary bacterial infections contributing to an estimated 50\% of COVID-19 deaths \citep{Mojica2022-xg, Chedid2021-eo}. Thus, the search for inhibitors of these enzymes is of considerable interest. To provide proof-of-concept of the CRowS approach, we performed four physical experiments for the inhibition of Verona Integron-encoded Metallo-$\beta$-lactamase 2 (VIM-2). Each of these experiments consisted of 1 known inhibitor and 30 known inert compounds, under a variety of conditions (Figure \ref{fig:expts}). The first experiment included no row constraints and an n/k ratio near 1. The other three experiments were more difficult in various ways, whether a smaller n/k ratio (Experiment 2) or row constraints (Experiments 3 and 4). The designs for each experiment were constructed using the procedure described in Section \ref{sec:r-cSSD}. The compounds were arrayed into normalized pools implied by the designs such that each compound was at a concentration of 30 $\mu$M. We transferred the compounds into the pools using an Eppendorf epMotion 5073m liquid handing workstation. Enzymatic assays were carried out in microplates at 25$^{\circ}$C and results were determined by reading the absorbance at 495 nm in a Molecular Devices SpectraMax iD5 plate reader. Additional details are in Supplementary Material A, while the list of compounds, the four designs, and the experimental results are provided in Supplementary Material D.

Figure \ref{fig:expts} shows the Lasso estimate profile plots for the four experiments. In the first three experiments, L-captopril is clearly identified as an inhibitor, though in Experiments 1 and 3 they also identify a possible false positive, Adenosine-5'-triphosphate disodium salt hydrate (abbreviated A-5'-dsh). The identification of 1 false positive out of 30 inert compounds is in keeping with a false positive rate of around 3\%. Note that in Experiment 2 we do not consider 1,6-Hexanediol (abbreviated 1,6-Hxnd) as a false positive because its potential effect is not inhibitory. We see that even in experiment 3, in which the design limits the number of compounds per well to 10, we clearly identify the known inhibitor. On the other hand, in the fourth experiment, where only five compounds per well are allowed, we fail to identify the known inhibitor though we also identify no false positives.

\section{Discussion} \label{sec:discussion}

In this paper, we have developed and studied a new class of supersaturated designs, which we call CRowS designs,  motivated by applications in high-throughput screening and in particular to the search for enzymes that cause antibiotic resistance. In these designs, the number of compounds per well is constrained and we use the $UE(s^2)$ criterion to guide their construction. We present an efficient exchange algorithm to construct them, present some initial lower bounds on the criterion value, study the effect of the row constraint upon the designs, show that they compare quite favorably to common competitors, and demonstrate that they detect a metallo-$\beta$-lactamase inhibitor in real experiments. Thus, our work presents an innovative statistical design methodology inspired by an important scientific problem, along with the suggestion to use a statistical modeling approach to the analysis of the designs. In addition, these designs are larger than have typically been studied, and our methodology can scale to the hundreds or thousands of factors.

Though our approach represents a clear improvement over traditional HTS methodology, it also has several limitations. First, while our design construction procedure is fast for the sizes of designs reported in this paper, for designs with $n$ and $k$ both in the thousands (e.g. an experiment on a 1{,}536-well plate) may require on the order of days to produce an approximately optimal design. Also, this methodology should save on the number of required plates compared to one-compound-one-well, but may require more liquid handling since each well includes $c$ different compounds. This, of course, must be weighed against the improved hit rates that pooling using our method could provide. 

We also note that depending on the application, statistical or chemical interactions could be a concern. We briefly investigated a setting in which  statistical interactions are sparse (this could arise, for instance, if interactions only occur between compounds that are both hits), by repeating the simulations described in Section~\ref{sec:sim} where the response was generated using the model in \eqref{eq:model} but the $\betavec$ included two active factors and a single active interaction, half the size of the active main effects. We considered both synergistic and antagonistic interactions that obey strong (interaction only appears when both parents are active) or weak (interaction only includes one active parent) heredity. In Section 7 of Supplementary Material A, we find that synergistic interactions degrade the ability of CRowS to detect hit compounds, while antagonistic interactions improve the results. We note that \citet{singh2023selection} has recently proposed a method to estimate interactions in SSD's, and this could provide a possible solution to the problem of sparse interactions. There is also the possibility of ``promiscuous inhibitors'' \citep{feng2006synergy}, in which compounds may react to each other in unpredictable ways. This problem may be application-specific, or specific to particular chemical groups, and warrants additional exploration in future studies seeking to expand the number of compounds that can be simultaneously tested.

Our work raises a number of interesting follow-up questions as well. On the application side, can the methodology be used in larger-scale settings to identify potential hits? In large-scale screens, can our proposed methodology be adjusted to largely eliminate false positives in large effect environments, or be used as a secondary screening tool to eliminate false positives that are identified in the initial screen? Theoretically, can the lower bounds be tightened and the reliance on the assumption of tight constraints be relaxed? More fundamentally, is it possible to more rigorously characterize conditions that will guarantee tight row-constraints, rather than the empirical observation we've made that row-constraints are tight until $c$ is larger than 40\%-45\% of $k$? We leave these questions for future work. In general, our work opens up new applications and development for SSDs. These CRowS designs can be applied to a broad class of biological and chemical screening problems, and the size of the designs are much larger than those that have been typically studied in the SSD literature. We note that \citet{eskridge2004large} study large SSDs, and even give a nod to HTS as an application, but they don't provide designs for $n>24$ nor do they consider row constraints or provide a real example.

\subsection*{Supplementary Material}

\begin{itemize}
    \item[A] Supplementary document which includes proofs, additional results, and details regarding the real experiments.
    \item[B] Matlab code to construct CRowS designs.
    \item[C] Designs from the paper, excepting the designs for the real experiments.
    \item[D] Designs, data, and analysis of the real experiments.
\end{itemize}

\subsection*{Acknowledgements}

This work was funded in part by Miami University's Faculty Research Committee (BJS) and by the National Institutes of Health through grant GM128595 (RCP). We also thank Jens Mueller from Miami's Research Computing Support for his computational assistance. 

\subsection*{Disclosure and Data Availability Statement}

The authors report there are no competing interests to declare. The data underlying this article are available in its online supplementary material.

\bibliographystyle{asa} 
\bibliography{main-Biometrics.bib}

\section{Supplementary Material}

\subsection{Proof of Proposition 1}

\begin{proof}
Because $\tilde s_{jl}-s_{jl} = \sum_{h=1}^n(\tilde x_{hj}\tilde
x_{hl}-x_{hj}x_{hl})$ with $\tilde X$ and $X$ differing only in row $i$, we
see that
\[\tilde s_{jl}-s_{jl}=\tilde x_{ij}\tilde x_{il}-x_{ij}x_{il}
  =\begin{cases}
    -2x_{ij}x_{il},
       & \tilde x_{ij}\not=x_{ij} \text{ or } \tilde x_{il}\not=x_{il}
         \text{ but not both},\\
    0, &\text{otherwise.}
    \end{cases}
\]
Thus
\begin{align*}
 \tilde s_{jl}^2-s_{jl}^2
  &=\begin{cases}
    (s_{jl}-2x_{ij}x_{il})^2-s_{jl}^2,
       & \tilde x_{ij}\not=x_{ij} \text{ or } \tilde x_{il}\not=x_{il}
         \text{ but not both},\\
    0, &\text{otherwise}
    \end{cases}\\
  &=\begin{cases}
    4-4x_{ij}s_{jl}x_{il},
       & \tilde x_{ij}\not=x_{ij} \text{ or } \tilde x_{il}\not=x_{il}
         \text{ but not both},\\
    0, &\text{otherwise.}
    \end{cases}
\end{align*}
This allows us to write
\begin{align*}
 Q(\tilde X)-Q(X)
  &=\sum_{j=0}^k\sum_{l=0}^k (\tilde s_{jl}^2-s_{jl}^2)
   =\sum_{j\in J}\sum_{l\in J} 0 + \sum_{j\not\in J}\sum_{l\not\in J} 0
    + 2\sum_{j\in J}\sum_{l\not\in J}
      (4-4x_{ij}s_{jl}x_{il})
      \\
  &=8|J|(k+1-|J|)
    -8\sum_{j\in J}\bigg[x_{ij}\sum_{l\not\in J}s_{jl}x_{il}\bigg],
\end{align*}
as claimed.
\end{proof}

\subsection{Proof of Theorem 1}
To prove this result, we adapt arguments from Liu and Hickernell (2002). The equation numbers (7)--(10) below refer to those in the statement of the theorem in the main article.

\begin{proof}
Equation (8) is immediate. To obtain (7), (9) and
(10) we first prove these identities:
\begin{gather}
  \|X\mathbf1_k\|_2^2 + \sum_{l=1}^k\sum_{j=1}^k\|X_{:,l}-X_{:,j}\|_1 = nk^2,
   \tag{S.1}\label{eq:rowsum+coldiff} \\
  \|X'\mathbf1_n\|_2^2 + \sum_{i=1}^n\sum_{m=1}^n\|X_{i,:}-X_{m,:}\|_1 = n^2k,
   \tag{S.2}\label{eq:colsum+rowdiff}\\
  \tr(X'XX'X)
  = -n^2k^2 + 2n\|X\mathbf{1}_k\|_2^2
     + \sum_{l=1}^k\sum_{j=1}^k\|X_{:,l}-X_{:,j}\|_1^2.
    \tag{S.3} \label{eq:trace-rowsumcoldiff}
\end{gather}
Because $ab + |a-b| = 1$ for any values $a,b \in \{\pm1\}$, we may write
\begin{align*}
    \|X\mathbf1_k\|_2^2 + \sum_{l=1}^k\sum_{j=1}^k\|X_{:,l}-X_{:,j}\|_1
   &=\sum_{i=1}^n\left[\sum_{l=1}^kx_{il}\right]^2
     + \sum_{l=1}^k\sum_{j=1}^k\sum_{i=1}^n|x_{il}-x_{ij}|\\
   &=\sum_{i=1}^n\sum_{l=1}^k\sum_{j=1}^kx_{il}x_{ij}
     + \sum_{i=1}^n\sum_{l=1}^k\sum_{j=1}^k|x_{il}-x_{ij}|
    =\sum_{i=1}^n\sum_{l=1}^k\sum_{j=1}^k1
\end{align*}
to obtain (\ref{eq:rowsum+coldiff}). The proof of (\ref{eq:colsum+rowdiff})
is analogous. Similarly, we see that
\begin{align*}
 \tr(X'XX'X)
  &= \sum_{l=1}^k\sum_{j=1}^k \left[ \sum_{i=1}^nx_{il}x_{ij} \right]^2 \\
  &= \sum_{l=1}^k\sum_{j=1}^k \left[ \sum_{i=1}^n(1-|x_{il}-x_{ij}|) \right]^2
   = \sum_{l=1}^k\sum_{j=1}^k \Big[ n-\|X_{:,l}-X_{:,j}\|_1 \Big]^2.
\end{align*}
Expanding the square in this last expression and then substituting
(\ref{eq:rowsum+coldiff}) yields
\begin{align*}
  \tr(X'XX'X)
  &=\sum_{l=1}^k\sum_{j=1}^k
    \Big[ n^2-2n\|X_{:,l}-X_{:,j}\|_1+\|X_{:,l}-X_{:,j}\|_1^2 \Big] \\
  &= n^2k^2-2n\left[nk^2-\|X\mathbf1_k\|_2^2\right]
    + \sum_{l=1}^k\sum_{j=1}^k\|X_{:,l}-X_{:,j}\|_1^2,
\end{align*}
giving us (\ref{eq:trace-rowsumcoldiff}). The expression in
(7) is now obtained by substituting the value of
$\tr(X'XX'X)$ from (\ref{eq:trace-rowsumcoldiff}) into $Q(X) = n^2 +
2\|X'\mathbf{1}_n\|_2^2 + \tr(X'XX'X)$. Next we make the following general
claim:
\begin{quote}
If $\alpha\in\mathbb{Z}$, $q\in\mathbb{N}$, and $f:\mathbb{R}\to
\mathbb{R}\cup\{\infty\}$ is convex, then
\[ \min_{\textstyle y\in\mathbb{Z}^q}
   \left\{ \sum_{i=1}^q f(y_i) \right.\left|\, \sum_{i=1}^qy_i=\alpha \right\}
   = q_0f(\beta) + (q-q_0)f(\beta+1),
\]
where $\beta:=\lfloor\alpha/q\rfloor$ corresponds to dividing the dimension
$q$ into the constraint value $\alpha$ with remainder $q-q_0$.
\end{quote}
To derive the bounds (9) and (10)
from this claim, we write $X = 2U - \mathbf1_n \mathbf1_k'$ for
$U\in\{0,1\}^{n\times k}$. The bound (9) is obtained by
applying the claim with the choices $\alpha=nc$, $q=k$, $f(t)=(n-2t)^2$ and
$y=U'\mathbf{1}_n$. To obtain the bound (10), apply the
claim with $q=k^2-k$ and $f(t)=4t^2$, taking the entries of $y$ to be the
values of $\|U_{:,l}-U_{:,j}\|_1$ for $l\not=j$ so that $\alpha=2nc(k-c)$ by
(8) and (\ref{eq:rowsum+coldiff}).

To complete the proof, we justify the claim. If $y\in\mathbb{Z}^q$ satisfies
$\sum_iy_i=\alpha$ and $|y_i-y_j|\leq1$ for all $i,j$, then by integrality
the distribution of entries in $y$ must agree with those given by the vector
$y^*$ having entries
\[ y^*_m =
  \begin{cases} \beta,&m\leq q_0,\\ \beta+1,&m>q_0. \end{cases}
\]
Therefore, it suffices to show that some optimal choice has $|y_i-y_j|\leq1$
for all $i,j$. Consider $y$ with $\sum_ly_{l\vrule height8pt width0pt depth
0pt}=\alpha$ and $y_i-y_j>1$ for some coordinate indices $i,j$.  Define $g(y)  := \sum_i f(y_i)$ and let $z\in\mathbb{Z}^q$ be a copy of $y$ for which the entries $y_i$ and $y_j$ are
swapped. Also, define $\lambda = 1/(y_i-y_j) \in (0,1]$ and let $\tilde y = \lambda
z + (1-\lambda)y$. Then $\sum_l\tilde y_{l\vrule height8pt width0pt depth
0pt}=\alpha$, $g(\tilde y) \leq \lambda g(z) + (1-\lambda) g(y) = g(y)$, and
\[ \tilde y_l =
    \begin{cases}
      y_j+1,& l=j,\\ y_i-1,& l=i,\\ y_l,&\text{ else.}
    \end{cases}
\]
Thus $\tilde y\in\mathbb{Z}^q$ with $g(\tilde y) \leq g(y)$, establishing the
claim.
\end{proof}

\subsection{Empirical Investigation of Row Constraints (additional scenarios)}

This section expands the empirical investigation of Section 3.2 in the main article by considering the following $(n,k)$ combinations: $(30,31)$, $(96,192)$, and $(96,292)$ (Figures 1-6). As in the example in Section 3.2, row-constraint slack begins to appear at between 40\% and 45\% of $k$. Furthermore, once slack starts appearing, measures of design quality appear close to an asymptote.

\begin{figure}[H]
\centering
\includegraphics[width=100mm]{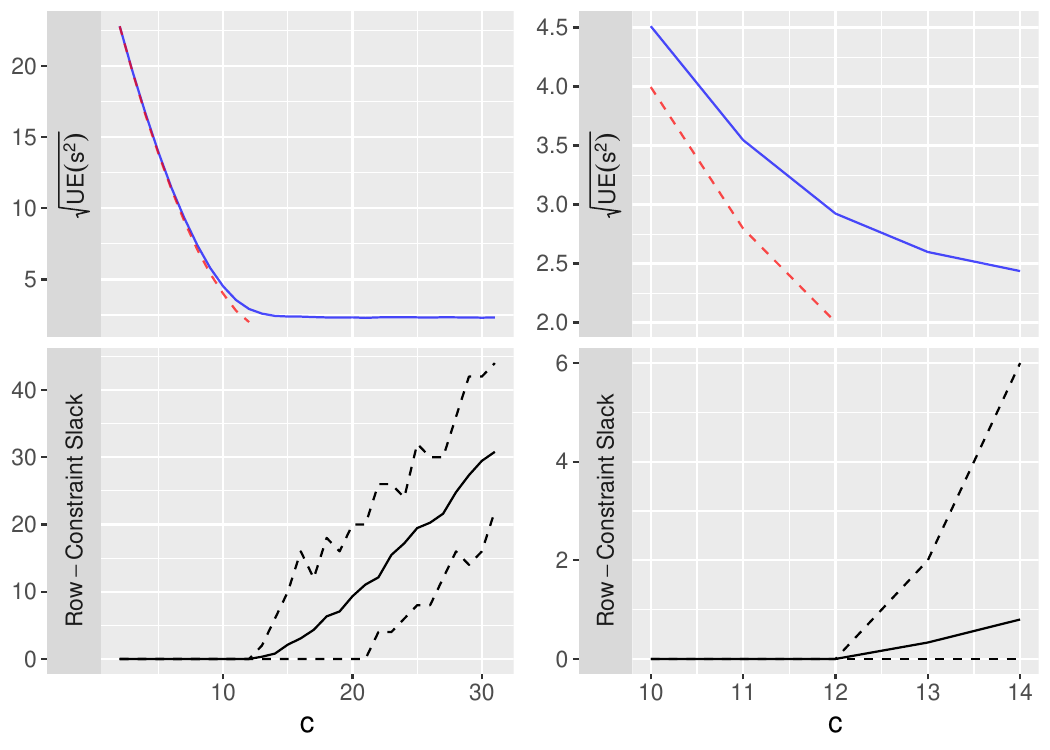}
\caption{\small Characteristics of CRowS designs with $n=30$, $k=31$ and $c \in \{2,3, \ldots, 31\}$. Plots are a function of $c$, the number of $+1$'s allowed in each row of design. Top left: $UE(s^2)$ value for each design (blue) along with the bounds from Section 3.1 (for all designs with no row-constraint slack). Bottom left: $(2c-k)-\sum_{j=1}^{k}x_{ij}$; the solid line is the average slack, while the lower and upper dashed lines are the minimum and maximum slack. Top right: The same as the top left plot, for $10 \leq c \leq 14$. Bottom right: The same as the bottom left plot, for $10 \leq c \leq 14$.}
\label{fig:n30_k31_cs}
\end{figure}

\begin{figure}[H]
\centering
\includegraphics[width=100mm]{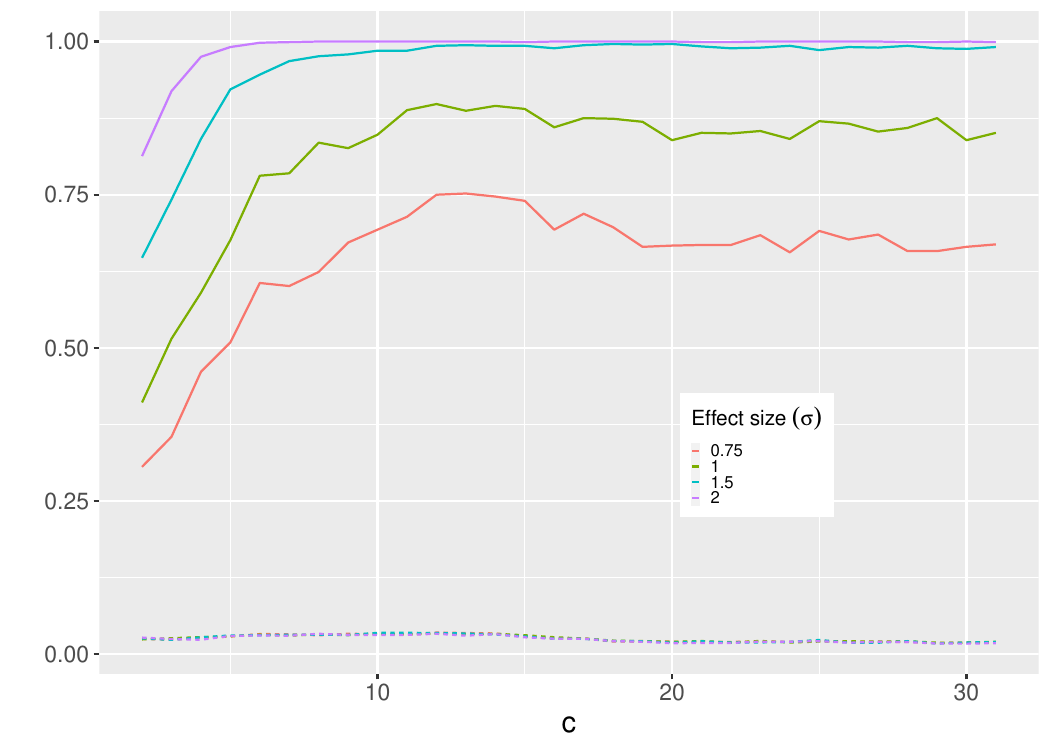}
\caption{\small Simulation results for CRowS designs with $n=30$, $k=31$ and $c \in \{2,3, \ldots, 31\}$, for four effect sizes, assuming one active factor; the solid lines are true positive rates while the dashed lines are false positive rates.}
\label{fig:n30_k31_cs}
\end{figure}

\begin{figure}[H]
\centering
\includegraphics[width=100mm]{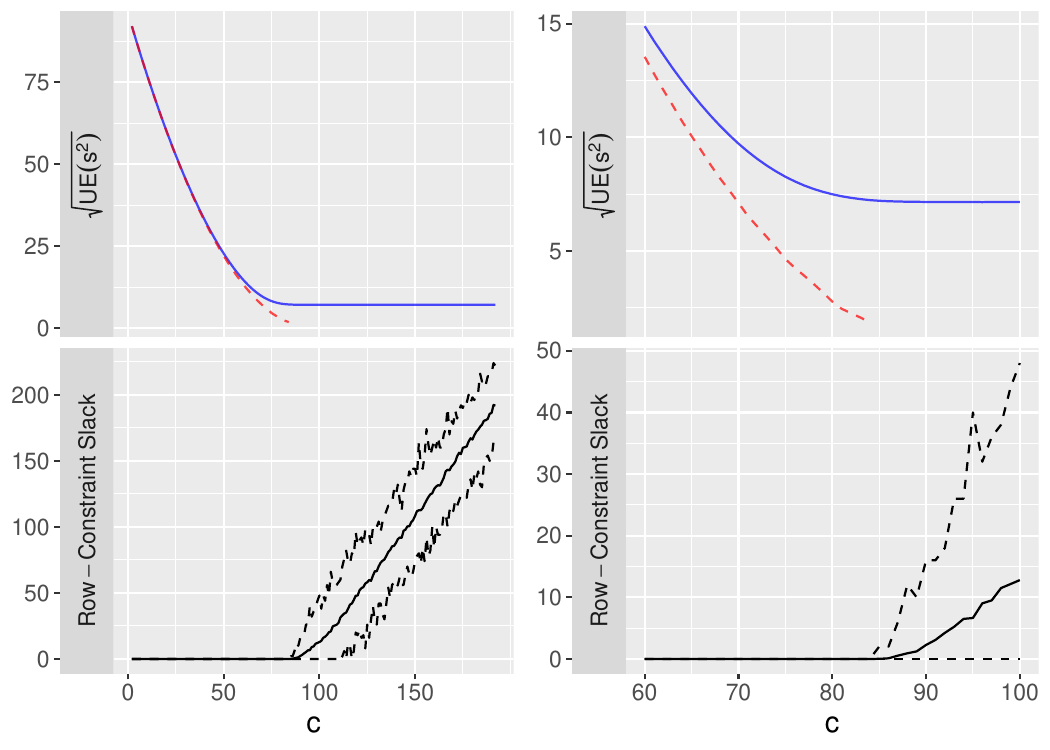}
\caption{\small Characteristics of CRowS designs with $n=96$, $k=192$ and $c \in \{2,3, \ldots, 192\}$. Plots are a function of $c$, the number of $+1$'s allowed in each row of design. Top left: $UE(s^2)$ value for each design (blue) along with the bounds from Section 3.1 (for all designs with no row-constraint slack). Bottom left: $(2c-k)-\sum_{j=1}^{k}x_{ij}$; the solid line is the average slack, while the lower and upper dashed lines are the minimum and maximum slack. Top right: The same as the top left plot, for $60 \leq c \leq 100$. Bottom right: The same as the bottom left plot, for $60 \leq c \leq 100$.}
\label{fig:n96_k192_cs}
\end{figure}

\begin{figure}[H]
\centering
\includegraphics[width=100mm]{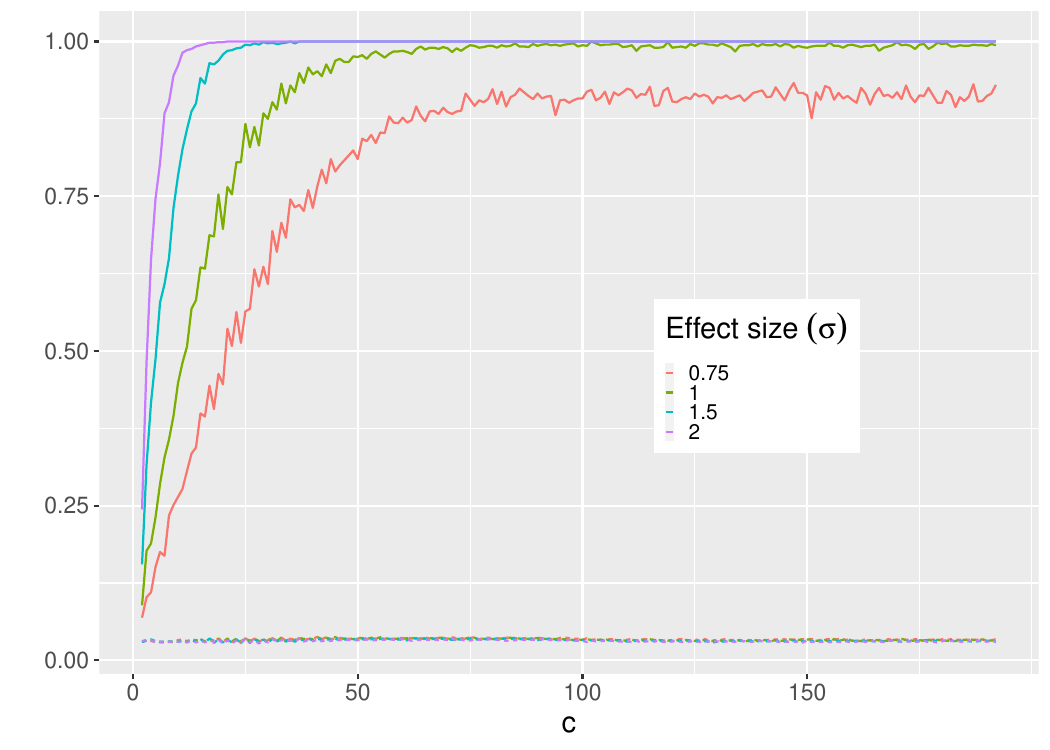}
\caption{\small Simulation results for CRowS design with $n=96$, $k=192$ and $c \in \{2,3, \ldots, 192\}$, for four effect sizes, assuming one active factor; the solid lines are True Positive Rates while the dashed lines are False Positive Rates.}
\label{fig:n96_k192_cs}
\end{figure}

\begin{figure}[H]
\centering
\includegraphics[width=100mm]{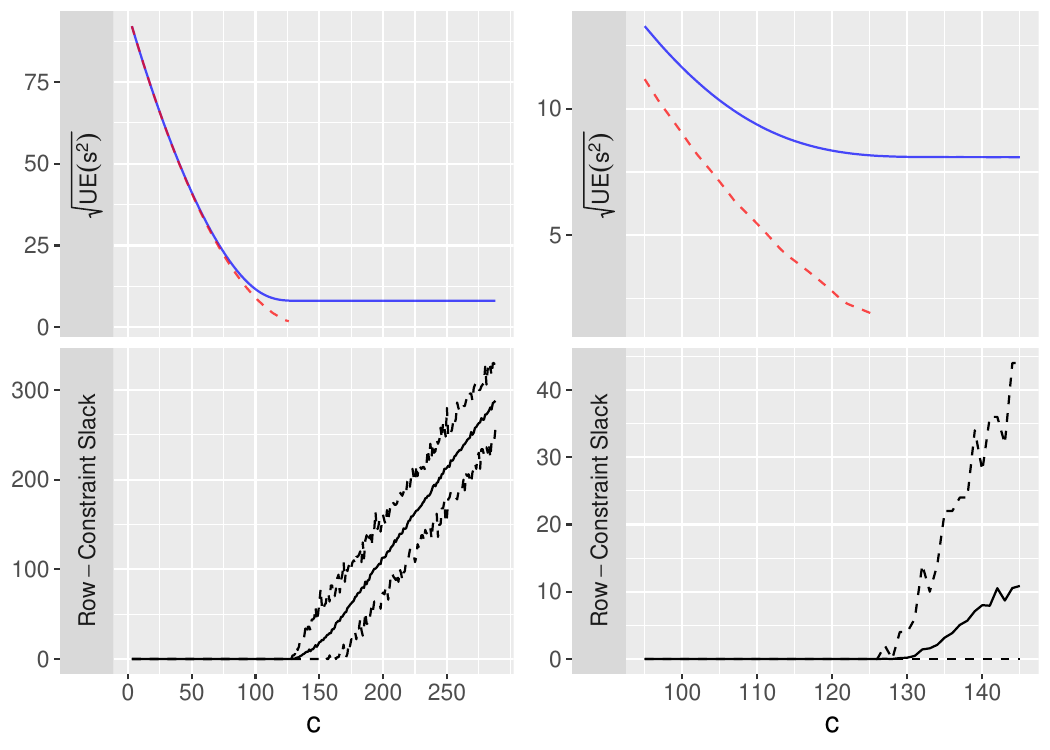}
\caption{\small Characteristics of CRowS designs with $n=96$, $k=288$ and $c \in \{3, 4, \ldots, 192\}$. Plots are a function of $c$, the number of $+1$'s allowed in each row of design. Top left: $UE(s^2)$ value for each design (blue) along with the bounds from Section 3.1 (for all designs with no row-constraint slack). Bottom left: $(2c-k)-\sum_{j=1}^{k}x_{ij}$; the solid line is the average slack, while the lower and upper dashed lines are the minimum and maximum slack. Top right: The same as the top left plot, for $95 \leq c \leq 145$. Bottom right: The same as the bottom left plot, for $95 \leq c \leq 145$.}
\label{fig:n96_k288_cs}
\end{figure}

\begin{figure}[H]
\centering
\includegraphics[width=100mm]{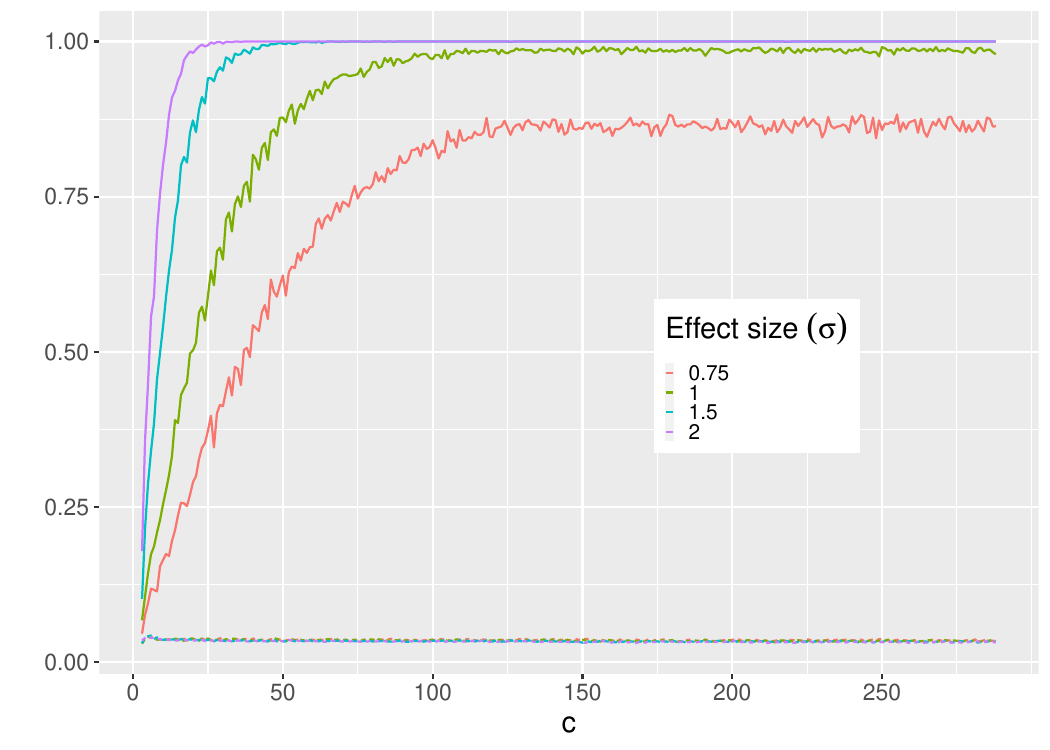}
\caption{\small Simulation results for CRowS designs with $n=96$, $k=288$ and $c \in \{3, 4, \ldots, 288\}$, for four effect sizes, assuming one active factor; the solid lines are true positive rates while the dashed lines are false positive rates.}
\label{fig:n96_k288_cs}
\end{figure}

\subsection{Description of poolHiTS}

Details can be found in \citet{kainkaryam2008poolhits}, but here we briefly describe the basic poolHiTS methodology. The user specifies the total number of compounds $k$ and the maximum number of compounds per well $c$, along with the anticipated maximum number of active compounds and an overall error rate. This allows the construction of a design with $a$ layers, each layer consisting of $q$ pools, where the parameters $a$ and $q$ are determined by the procedure. This results in a final STD with $n=aq$ wells, which highlights its lack of flexibility to choose the number of wells for a given $k$ and $c$. The decoding algorithm requires a binary response vector; thus, to account for uncertainty associated with such a binarization, we specify a threshold, much like that for OCOW, which we can use to label each well.  
Once the pools have been labeled, the decoding algorithm proceeds by eliminating compounds if they appear in at least $E+1$ inert wells, and tagging compounds as active if they aren't labeled as inert and appear in at least $E+1$ positive pools. Here $E$ is the number of errors allowed in the analysis. We note that it is possible for some compounds to fall into a third inconclusive category, and in this case we label these as active since further testing would be required to eliminate them.

\subsection{Method Comparisons (unknown $\mu$ and $\sigma^{2}$)}

For convenience and simplicity, in the paper we assumed that both $\mu$ and $\sigma^{2}$ are known in the simulations of Section 4, where $\mu$ is the background mean response value and $\sigma^{2}$ is the error variance of model (1) in the paper. However, these assumptions are unnecessary. In this section we show that the basic simulation results hold up even if we assume that $\mu$ and $\sigma^{2}$ are unknown. That is, we provide versions of OCOW, poolHiTS, and CRowS that allow $\mu$ and/or $\sigma^{2}$ to be estimated via a pilot experiment, and then show a version of the paper's Figure 2 using these methods. For the simulations in this section, as an example, we assume that the pilot experiment is $n=12$ independent runs of known, non-hit compounds.

\subsubsection{One-Compound-One-Well: Unknown $\mu$ and $\sigma^{2}$} \label{sec:OCOW_unknown_sigmasq}

The One-Compound-One-Well (OCOW) approach is described in Section 4.1.1 of the paper, and can be briefly summarized as: deposit a single biological entity in each well, measure the inhibition of each well, and declare as potentially active any compound whose measurement exceeds a specified threshold. If $\mu$ and $\sigma^{2}$ are known, we provide a reasonable threshold in Section 4.1.1. If we don't know these parameters, we can use the method of \citet{lenth1989quick} to estimate $\sigma^{2}$ and use a pilot study to estimate $\mu$, thus establishing a threshold beyond which compounds are declared potential hits. Lenth's method was originally used to provide robust estimates of uncertainty for factorial effects in two-level factorial experiments fit to saturated models, but the basic idea applies to any set of normally distributed random variables where most have the same mean but a few have larger means (in absolute value). 

Let $y_1, y_2, \ldots, y_k$ be the measurements for $k$ compounds studied via OCOW, and $c_{1}, c_{2}, \ldots, c_{k}$ the centered versions of the measurements where $c_{i}=y_{i}-\mu$. Then, the $c_{i}$'s are assumed to be independent realizations of $N(\mu_{i},\sigma^{2})$. In our application setting, most compounds are inert while a few may be hits. Let $\mathcal{I}$ be the set of inert compounds and $\mathcal{H}$ be the set of hit compounds. Then, for $i \in \mathcal{I}$, $\mu_{i}=0$, while for $i \in \mathcal{H}$, $\mu_{i}>0$ (assuming WLOG positive effects). Lenth suggests the following to estimate $\sigma^{2}$: First, compute
\begin{align*}
    s_{0} = 1.5 \times \underset{i}{\text{median}} \; |c_{i}|,
\end{align*}
and then compute the pseudo standard error (PSE) as
\begin{align*}
    PSE = 1.5 \times \underset{\small |c_{i}|<2.5s_{0}}{ \text{median}} \; |c_{i}|.
\end{align*}
For reasons discussed in \citet{lenth1989quick}, this is a good estimate of $\sigma$ under sparsity assumptions. Since we are looking for positive effects, our threshold to declare a compound a hit is $t_{0.95,k/3} \times PSE$, where $k/3$ is also recommended by Lenth. We assume that during each simulation, and estimate of $\mu$ is drawn from $N(\mu, \sigma^{2}/12)$. This procedure can be used to analyze simulated data under an OCOW design, without assuming knowledge of $\mu$ and $\sigma^{2}$.

\subsubsection{poolHiTS: Unknown $\mu$ and $\sigma^{2}$} \label{sec:poolHiTS_sigmasq_unknown}

We described the poolHiTS procedure in Section 4.1.2 of the main article; more details can be found in \citet{kainkaryam2008poolhits}. An important part of the approach is determining whether a given well (containing multiple compounds) is a hit or not. In particular applications, there may be an existing test which can be used. In our setting, we use a similar threshold as with OCOW: declare a well as a hit if the response falls beyond $\mu + F^{-1}(0.96)\sigma$, where $F^{-1}$ is the inverse standard normal distribution function and the 96th percentile was chosen to try to balance true positive and false positive rates in the simulations in Section 4 of the paper. 

To relax the assumption that $\mu$ and $\sigma^{2}$ are known, we assume an OCOW pilot experiment is conducted with $12$ known inert compounds, from which the sample mean $\overline{X}$ and sample variance $S^{2}$ is computed. Then, the threshold is computed as $\overline{X} + F^{-1}(0.96)\sqrt{S^{2}}$. For our simulation, at each iteration we simulate $\overline{X} \sim N(\mu,\sigma^{2}/12)$ and $11 S^{2}/\sigma^{2} \sim \chi^{2}_{11}$, from which we obtain $\overline{X}$ and $S^{2}$ for the threshold.

\subsubsection{CRowS: Unknown $\mu$ and $\sigma^{2}$} \label{sec:SSD_unknown_sigmasq}

For our proposed procedure, knowledge of $\mu$ is not required. However, in the version of the procedure described in Section 4.1.3 of the document, we use knowledge of $\sigma^{2}$ in order to threshold within the Lasso procedure. Thus, we employ the same pilot experiment assumptions that we described above in Supplementary Section \ref{sec:poolHiTS_sigmasq_unknown}, which allows us to simulate with the relaxed assumption that $\sigma^{2}$ is not known but estimated by $S^2$.

\subsubsection{Simulation Results (Unknown $\mu$ and $\sigma^{2}$)}

Here we present results (analogous to those in Section 4.3 and Figure 3 in the main article) for the case in which $\mu$ and $\sigma^{2}$ are not assumed known. Besides the differences in the analysis and simulations described above in Sections \ref{sec:OCOW_unknown_sigmasq}-\ref{sec:SSD_unknown_sigmasq}, the simulation settings are the same.

The basic conclusions regarding the three methods are unchanged. The results from our proposed method and OCOW are similar compared to the versions with known $\mu$ and $\sigma^{2}$; the poolHiTS results are somewhat worse. Overall, it is still clear that our proposed method is preferred for the designs and settings considered here.

\begin{figure}[h]
    \centering
     \begin{subfigure}{0.6\textwidth}
         \includegraphics[width=\textwidth]{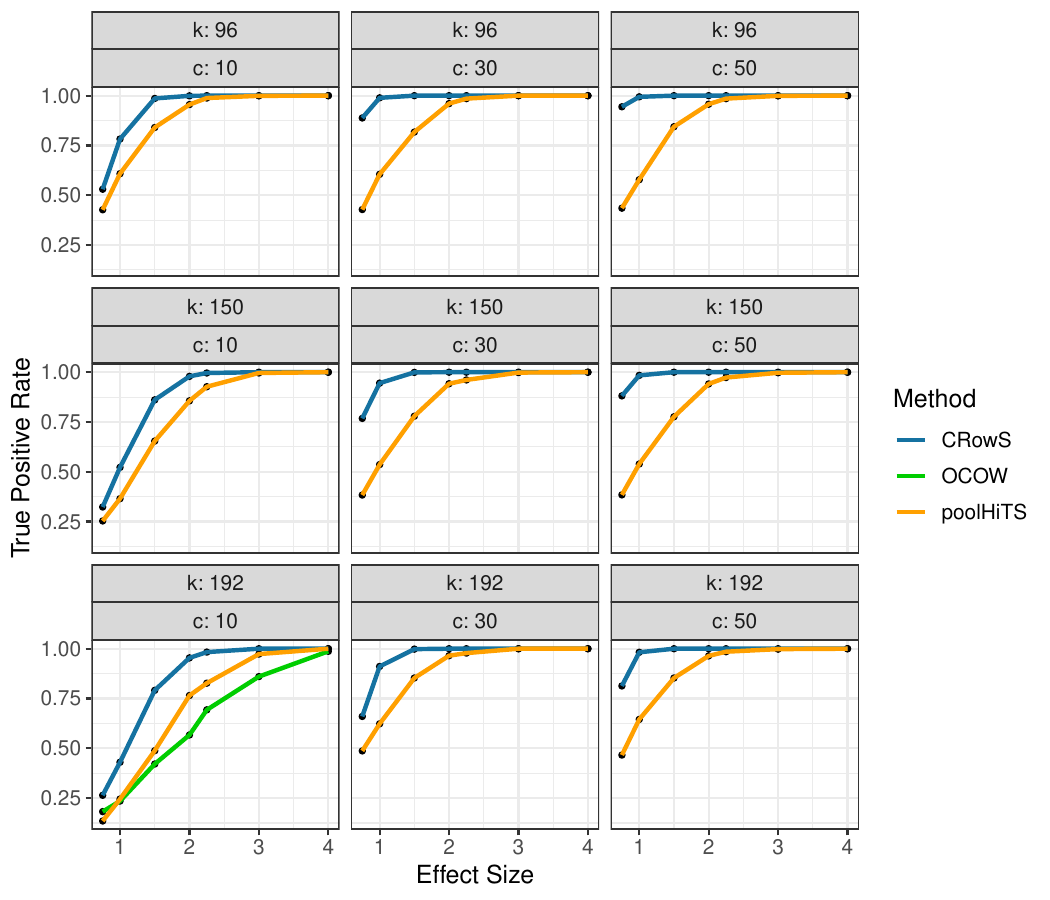}
         \caption{\footnotesize True positive rates for the three methods (unknown $\mu$ and $\sigma^{2}$).}
         \label{fig:comp_tpr}
     \end{subfigure}
     \hfill
     \begin{subfigure}{0.6\textwidth}
         \includegraphics[width=\textwidth]{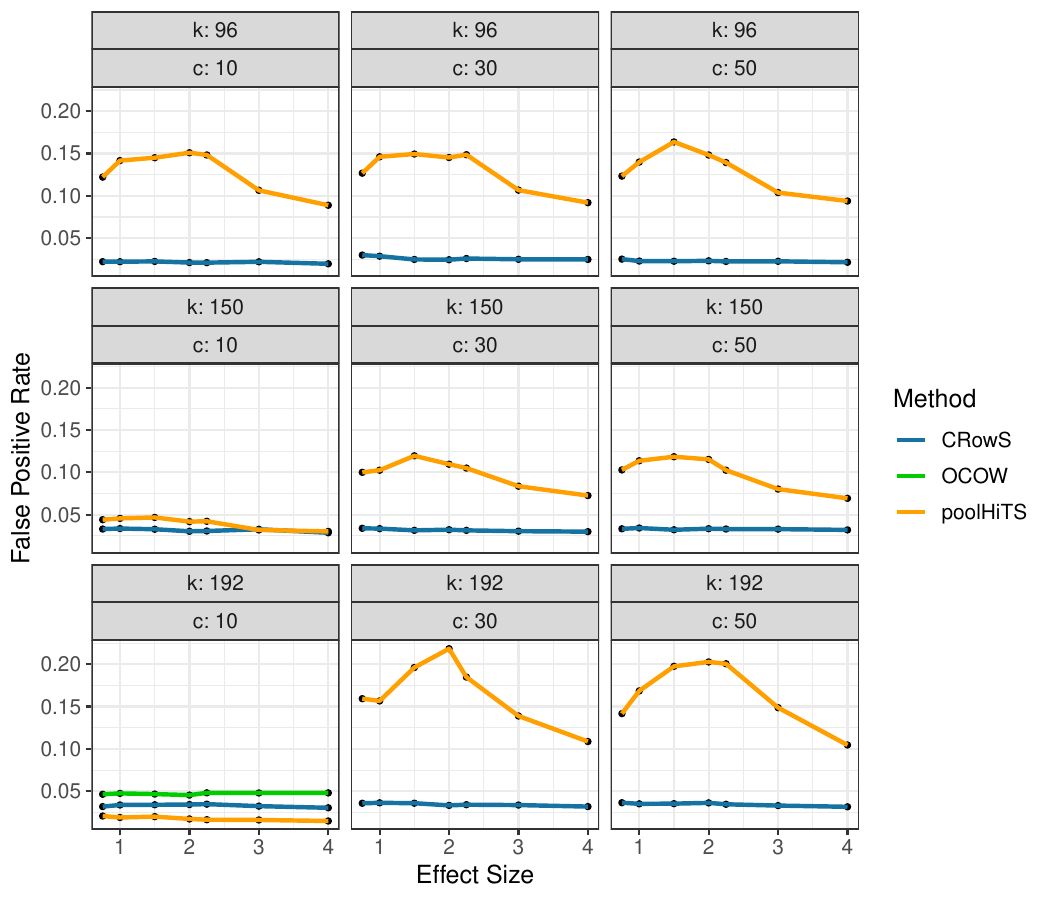}
         \caption{\footnotesize False positive rates for the three methods (unknown $\mu$ and $\sigma^{2}$).}
         \label{fig:comp_fpr}
     \end{subfigure}
     \caption{\small Analogous to Figure 2 in the main document, with relaxed assumptions about known $\mu$ and $\sigma^{2}$.}
     \label{fig:comp}
\end{figure}

\subsection{Real Experiments' Details}

\begin{figure}
\centering
\includegraphics[width=162mm]{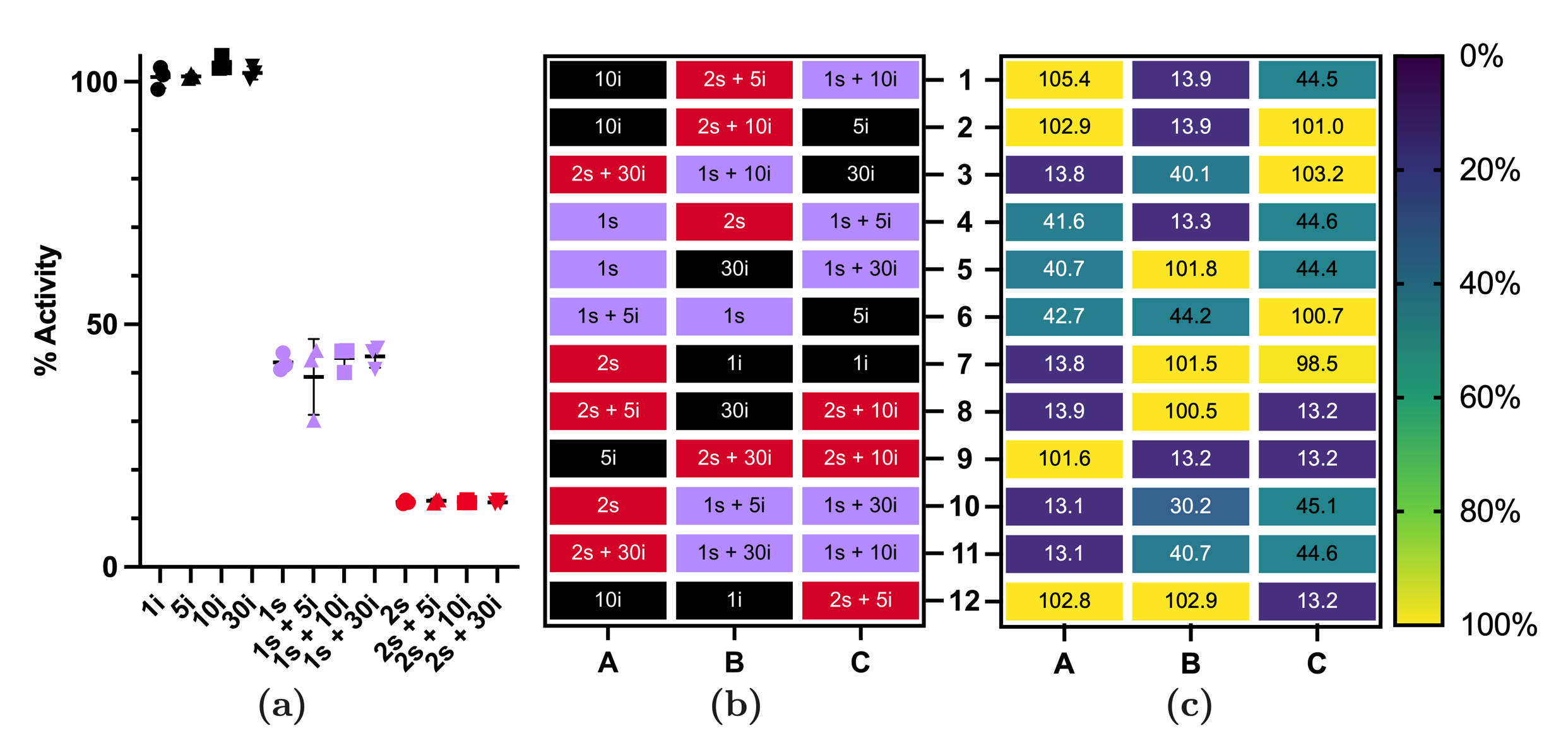}
\caption{Pilot screen exploring the effects of pool sizes. Each assay well contained 5.4 nM VIM-2, 50 mM HEPES, pH 7.0, 2 mM CHAPS, and 144 $\mu$M nitrocefin. Pooled screen compounds included 1 inert (1i), 5 inert (5i), 10 inert (10i), 30 inert (30i), 1 spiked (1s), 1 spiked + 5 inert (1s + 5i), 1 spiked + 10 inert (1s + 10i), 1 spiked + 30 inert (1s + 30i), 2 spiked (2s), 2 spiked + 5 inert (2s + 5i), 2 spiked + 10 inert (2s + 10i), and 2 spiked + 30 inert (2s + 30i). Wells with 1 spiked verified inhibitor compound feature 50 $\mu$M L-captopril. Wells with 2 spiked verified inhibitor compounds feature 50 $\mu$M L-captopril and 50 $\mu$M 2,3-dimercaprol. (a): The percent enzyme activity, relative to VIM-2 without any added screen compounds, is provided. Error bars indicate the standard deviation for 3 replicates. All wells with 1 inert, 1 spiked, or 2 spiked verified inhibitor compounds are drawn as circles. All wells containing 5 inert compounds are drawn as triangles. All wells containing 10 inert compounds are drawn as squares. All wells containing 30 inerts are drawn as upside down triangles. Wells with no spiked inhibitors are black, with 1 spiked inhibitor are purple, and 2 spiked inhibitors are red. (b): Map of the plate using random mapping of pools to assay wells using the same coloring as in (a). (c): Heat map of the percent enzyme activity for each well of the assay plate, relative to VIM-2 without any added screen compounds.}
\label{fig:pilot}
\end{figure}

Before conducting the experiments described in Section 5 of the main article, we conducted a pilot experiment (Figure \ref{fig:pilot}). This pilot included the 31 compounds from the experiments in Section 5, along with another known inhibitor, 2,3-dimercaprol. The pilot experiment clearly shows that adding the inert compounds did not affect the level of inhibition. It also shows small unit-to-unit variation and large effect sizes for the known inhibitors.

The 31 compounds studied in the four experiments in Section 5 are provided in \verb$CompoundMap.csv$, while the designs are contained in \verb$4Expts.csv$ and the responses are in \verb$4Expts_data.csv$. Note that the four experiments (Table 2 in the main article) were conducted on the same 96-well plate, which also include several control wells. Experiment 1 was conducted in Wells 1-26; Experiment 2 was conducted in Wells 29-44; Experiment 3 was conducted in Wells 47-70; and Experiment 4 was conducted in Wells 73-96. For the control wells (see \verb$4Expts.csv$):
\begin{itemize}
    \item Well 27 had only L-Captopril, the known inhibitor
    \item Well 28 had only a neutral liquid
    \item Well 45 had the 30 inert compounds
    \item Well 46 had only a neutral liquid
    \item Well 71 had the 30 inert compounds
    \item Well 72 had only L-Coptopril
\end{itemize}

\subsection{Simulation Results with Interactions}

\begin{figure}[H]
    \centering
     \begin{subfigure}{0.85\textwidth}
         \includegraphics[width=\textwidth]{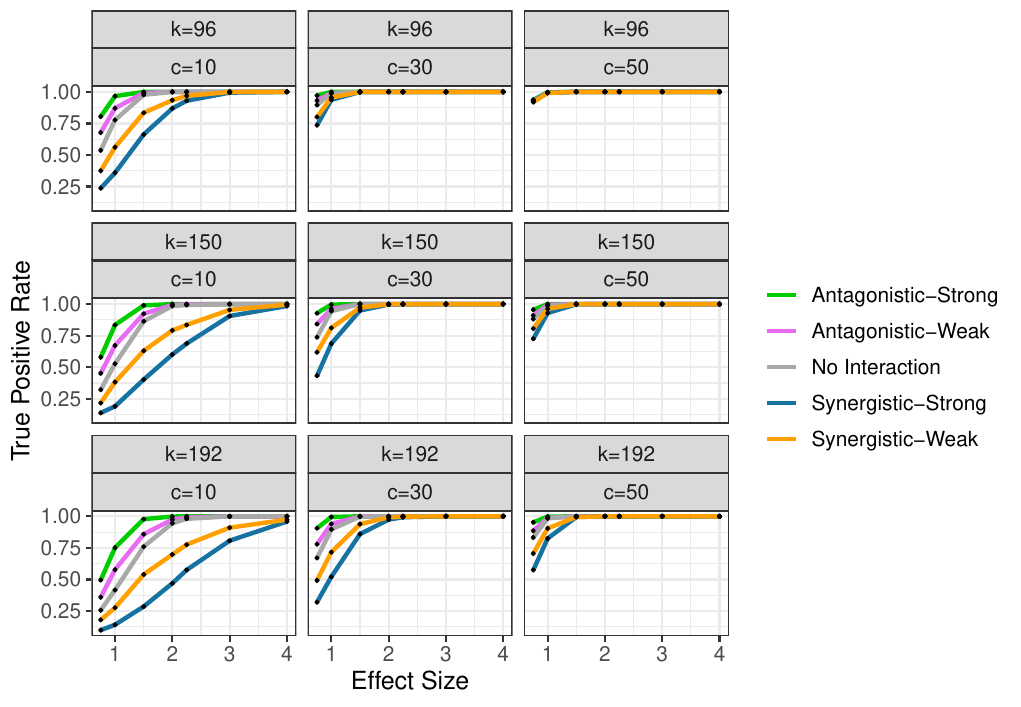}
         \caption{\small True Positive Rates.}
         \label{fig:int_comp_tpr}
     \end{subfigure}
     \hfill
     \begin{subfigure}{0.85\textwidth}
         \includegraphics[width=\textwidth]{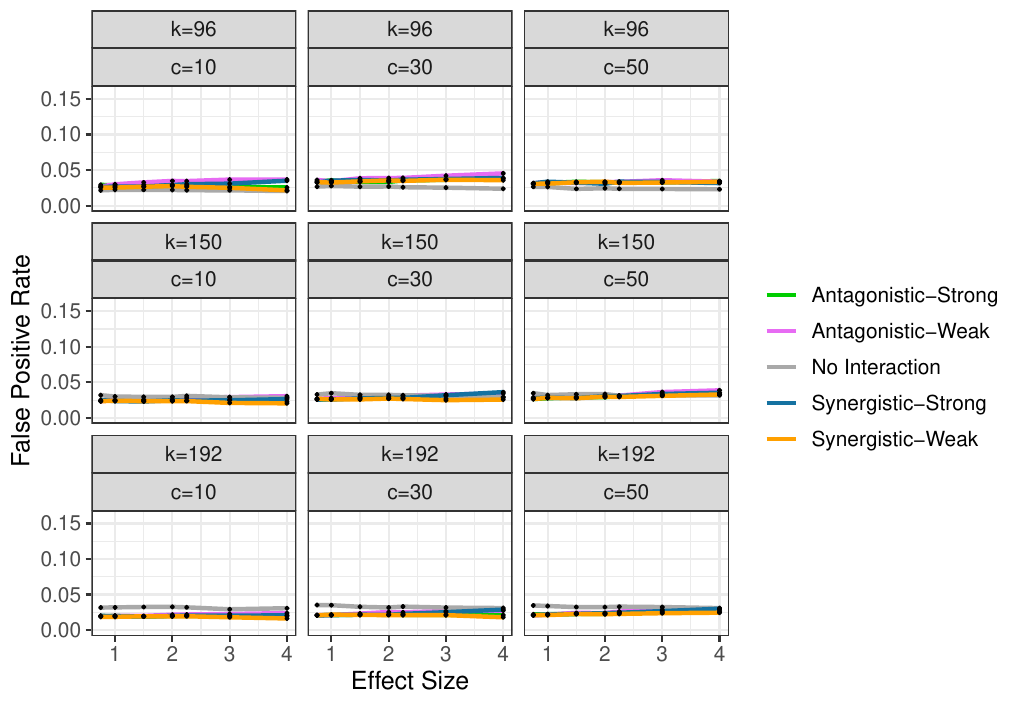}
         \caption{\small False Positive Rates.}
         \label{fig:int_comp_fpr}
     \end{subfigure}
     \caption{\small Evaluation of CRowS under various interaction scenarios. The true model includes two active compounds, and interactions as specified. The interaction effects are half the size of the main effects.}
     \label{fig:int_comp}
\end{figure}

Here we present the results of the simulation we describe in Section 6, which includes a true model with interaction. Figure~\ref{fig:int_comp_tpr} shows a decrease in the TPR for those models with smaller effects sizes and for designs with smaller values of $c$, i.e. $c=10$.  For larger effect sizes and designs with $c>10$ we observed no change in the TPR.  Figure~\ref{fig:int_comp_fpr} shows the FPR remains constant and is equivalent to the CRowS FPR shown in Figure~\ref{fig:comp_fpr}. 

\end{document}